\documentclass[useAMS,usenatbib]{mn2e}
\usepackage{deluxetable}
\usepackage{epsf}
\usepackage{amsmath}
\usepackage{subfigure}
\usepackage{url}

\newcommand{\kmsmpc}{\kms\;{\rm Mpc}^{-1}}
\newcommand{\lya}{Ly$\alpha$\ }
\newcommand{\hkpc}{h^{-1}{\rm kpc}}
\newcommand{\hmpc}{h^{-1}{\rm Mpc}}
\newcommand{\lcdm}{$\Lambda$CDM}
\newcommand{\kms}{\;{\rm km}\,{\rm s}^{-1}}

\newcommand{\msun}{M_{\odot}}
\newcommand{\mstar}{M_*}

\newcommand{\sfr}{\dot{M}_*}

\newcommand{\rhostar}{\rho_*}

\newcommand{\aap}{A\&A}
\newcommand{\apjs}{ApJS}
\newcommand{\apj}{ApJ}
\newcommand{\apjl}{ApJL}
\newcommand{\aj}{AJ}
\newcommand{\araa}{ARA\&A}
\newcommand{\mnras}{MNRAS}

\newcommand{\fesc}{f_{\mathrm{esc}}}

\begin{document}

\title{Smoothly-Rising Star Formation Histories During the Reionization Epoch}

\author[K.\ Finlator et al.]{
\parbox[t]{\textwidth}{\vspace{-1cm}
Kristian Finlator$^{1\dagger}$, Benjamin D.\ Oppenheimer$^2$, Romeel Dav\'e$^3$}
\\\\$^1$Department of Physics, University of California, Santa Barbara, CA
93106, USA
\\$^\dagger$Hubble Fellow
\\$^2$Leiden Observatory, Leiden University, PO Box 9513, 2300 RA Leiden,
Netherlands
\\$^3$Astronomy Department, University of Arizona, Tucson, AZ 85721, USA
\author[astronomy]{Kristian Finlator, Benjamin D.\ Oppenheimer, \& Romeel Dav\'e}
}

\maketitle

\begin{abstract}
Cosmological hydrodynamic simulations robustly predict that high-redshift 
galaxy star formation histories (SFHs) are smoothly-rising and vary 
with mass only by a scale factor.  We use our latest simulations to 
test whether this scenario can account for recent observations at $z\geq6$ 
from WFC3/IR, NICMOS, and IRAC.  Our simulations broadly reproduce the 
observed ultraviolet (UV) luminosity functions and stellar mass densities 
and their evolution at $z=6$--8, all of which are nontrivial tests of the
mean SFH.  In agreement with observations, simulated galaxies possess blue 
UV continua owing to young ages (50--150 Myr), low metallicities 
(0.1--0.5$Z_\odot$), and low dust columns ($E(B-V) \leq 0.05$).  Our 
predicted Balmer breaks at $z=7$, while significant, are $\approx0.5$ 
magnitudes weaker than observed even after accounting for nebular line 
emission, suggesting observational systematic errors and/or numerical 
resolution limitations.  Observations imply a near-unity slope in the 
stellar mass--star formation rate relation at all $z=6$--8, confirming the 
prediction that SFH shapes are invariant.  Dust extinction suppresses the 
UV luminosity density by a factor of 2--3, with suppression increasing 
modestly to later times owing to increasing metallicities.  Current 
surveys detect the majority of galaxies with stellar masses exceeding 
$10^9\msun$ and few galaxies less massive than $10^{8.5}\msun$, implying 
that they probe no more than the brightest $\approx30\%$ of the complete 
star formation and stellar mass densities at $z\geq6$.  Finally, we 
demonstrate that there is no conflict between smoothly-rising SFHs and 
recent clustering observations.  This is because momentum-driven outflows 
suppress star formation in low-mass halos such that the fraction of halos
hosting observable galaxies (the ``occupancy") is 0.2--0.4 even though the 
star formation duty cycle is unity.  This leads to many interesting predictions 
at $z\geq4$, among them that (1) optically-selected and UV-selected samples 
largely overlap; (2) few galaxies exhibit significantly suppressed specific 
star formation rates; and (3) occupancy is constant or increasing with 
decreasing luminosity.  These predictions are in tentative agreement with 
current observations, but further analysis of existing and upcoming data 
sets is required in order to test them more thoroughly.
\end{abstract}


\section{Introduction} \label{sec:intro}
In the era of precision cosmology, the study of galaxy evolution
is an initial value problem in which the goal is to characterize the 
processes that connect the initial conditions at 
$z\approx1090$~\citep[for example,][]{kom10} to galaxies, which had evidently 
begun forming by $z=10$~\citep[for example,][]{sta07,bou09c,yan09}.  The 
initial-value nature of the problem implies that detailed measurement 
of its earliest stages constitutes an immensely useful aid through 
its ability to anchor an understanding of the later stages.  Gas 
inflows, outflows, and other feedback processes impact high-redshift 
star formation histories (SFHs).  Hence observational inferences 
regarding the SFHs at high redshift constrain star formation and 
feedback both at early times and at subsequent epochs. 

Cosmological hydrodynamic simulations robustly predict smoothly-rising 
SFHs~\citep{fin07} with a scale-invariant shape at $z\geq6$.  Therefore, 
any comparison between simulation predictions and observations 
constitutes a (possibly indirect) test of this scenario.  We have 
previously demonstrated considerable success in using cosmological 
hydrodynamic simulations to interpret observations of high-redshift 
galaxies.  In~\citet{fin06}, we showed that simulations of a 
\lcdm~universe with star formation modulated by outflows with a 
constant wind speed and mass loading factor broadly reproduce the 
observed rest-frame UV luminosity function (LF) at $z\sim4$.  We 
confirmed previous numerical predictions~\citep[for example,][]{dav00,wei02} 
that stellar mass ($\mstar$) and star formation rate (SFR) correlate, 
finding a small scatter and a slope near unity.  A qualititatively 
similar correlation has now been observed at all redshifts out to 
$z=7$~\citep{bri04,elb07,noe07a,dad07,sal07,sch07,pan09,lab09,bot09,oli10,mag10}, 
although observations at $z<2$ tend to suggest a somewhat flatter 
slope of 0.7--0.9 that is associated with galaxy 
downsizing~\citep{bri04,elb07,noe07a,dad07,sal07,sch07,bot09,mag10}.  This 
correlation has two implications: First, it implies that UV-selected 
galaxies are generically the most massive (star-forming) galaxies at 
a given redshift~\citep{nag04,fin06} unless the scatter is quite 
large~\citep{wei02}.  Second, a slope of unity supports a scenario in 
which SFHs have a scale-invariant shape such that, on average, galaxies' 
SFHs differ only by a scale factor.  The SFR-$\mstar$ relation therefore 
constitutes a key prediction of galaxy evolution 
models~\citep[for example,][]{bou09,dut09}.  We additionally predicted that 
high-redshift galaxies should exhibit pronounced Balmer breaks because
smooth SFHs naturally yield evolved stellar populations.  With the 
arrival of rest-frame optical constraints from IRAC, this prediction has 
now received dramatic confirmation~\citep[for example,][]{bra08,cha05,dow05,
dun07,ega05,eyl05,eyl07,lab06,lai07,lab09,sta09,ver07,yan05,pen09,zhe09}, 
although the question of whether emission lines could be mimicking the 
observed Balmer breaks remains unresolved~\citep{sch09,sch10}.  

In~\citet[hereafter DFO06]{dav06}, we tested a treatment for momentum-driven 
outflows from star-forming regions against the available constraints on 
the rest-frame UV and \lya LFs at $z\sim6$, finding reasonable agreement 
in both cases.  This result was subsequently extended by~\citet{bou07}, 
who found that our simulation's prediction for the evolution of the 
UV LF normalization from $z=8\rightarrow4$ was well-matched 
by observations.  This is a nontrivial test of our predicted SFHs 
because constant or decaying SFHs would predict a different evolution 
of the UV LF than is observed.  

Finally, in~\citet{fin07}, we showed that our simulations could account 
for the rest-frame UV-to-optical spectral energy distributions (SEDs) 
of 5 out of 6 observed galaxies between $z=5.5$--6.5.  We
demonstrated that smoothly-rising SFHs are equally as plausible as more 
popular constant or exponentially-decaying models.  We then used these
comparisons to explore the implied physical properties such as SFR, 
$\mstar$, and age, finding good agreement with previous constraints
although with significantly narrower posterior probability distributions 
owing to the tight priors imposed by the physical correlations that our 
simulations predict.

While these studies are encouraging, their support for our SFH scenario
is somewhat indirect because, until recently, observations at $z\geq6$ 
did not clearly imply a unique SFH.  For example, any galaxy evolution 
model that predicted scale-invariant SFH shapes would predict an 
SFR-$\mstar$ relation with a slope of unity.  Similarly, any SFH shape 
can be tuned to reproduce the observed UV LF at one epoch, and bursty
SFHs could be tuned to reproduce it at multiple epochs.  The fundamental 
difficulty is that broadband SEDs do not contain enough information to
constrain the underlying SFHs of individual objects, even when analyzed 
through detailed SED-fitting studies~\citep[for example,][]{sha01,pap01}.

Since the completion of these previous works, a number of authors 
have used new data from the HST WFC3/IR and NICMOS cameras as well 
as Spitzer/IRAC to increase the number of objects at $z\geq6$ with 
published rest-frame UV to optical constraints 
\citep{bou09b,bou10a,bun09,cas09,fink09,gon09,lab09,lab10,mcl10,
oes10a,oes10b,ouc09,yan09}.  Using these new samples, we may now 
ask whether the \emph{mean} SFH of reionization-epoch galaxies is 
constrained even if \emph{individual} SFHs are not.
For example,~\citet{sta09} recently pointed out that the observed 
lack of evolution in the specific star formation rate (SSFR) of 
UV-selected galaxies from $z=$4--6~\citep[see also][]{ove09,gon09} 
rules out the possibility that the majority of galaxies observed 
at $z=4$ have been forming stars at constant or declining rates
since $z\geq5$.  This is because constant or declining SFHs would
predict higher SSFRs at earlier times, in conflict with their
observations.  They noted that smoothly-rising SFHs could explain 
their observations since in this case galaxies would grow in 
$\mstar$ and SFR simultaneously.~~\citet{pap10} studied
the evolution of the UV luminosity and stellar mass at constant number
density and concluded that observations require smoothly-rising SFHs and
are consistent with a scale-invariant shape from $z=8\rightarrow 3$.  
Finally,~\citet{mar10} argued in favor of 
exponentially-rising over decaying SFHs as models for star-forming 
galaxies at $z\sim2$ for two reasons.  First, they found that rising 
SFHs provide better fits both to observations and to predictions from 
semi-analytic models of galaxy formation.  Second, rising SFHs 
naturally explain higher-redshift samples as the fainter progenitors 
of lower-redshift samples.

These suggestions exemplify how statistical samples at multiple epochs
can constrain the mean SFH even if the SFHs of individual galaxies are 
unconstrained.  In this work, we use this idea to build upon 
the results of~\citet{fin07} by asking whether smoothly-rising SFHs can 
explain the observed statistical properties of galaxies at $z=6$--8.  
Broadly, since all of our simulated high-redshift galaxies experience 
smoothly-rising SFHs, any comparison with observations can be viewed 
as an indirect test of these models.  In most cases, however, the 
constraining power of large samples at multiple epochs will render 
our comparisons direct tests of the smoothly-rising scenario.

We describe our simulations in Section~\ref{sec:sims}.  In
Section~\ref{sec:meanSFH}, we review the rising SFH scenario and 
summarize its predictions.  We compare our predicted UV LFs with 
observations in Section~\ref{sec:lfs}.  We perform detailed comparisons
between simulated and observed SEDs in Section~\ref{sec:colors}.
Additionally, we use our simulations to predict the completeness of 
current $z\geq6$ surveys.  In Section~\ref{sec:phys}, we study the 
predicted relationships between stellar mass, star formation rate, 
star formation history, and metallicity, comparing with observations 
where possible.  In Section~\ref{sec:occ}, we use our predicted halo 
occupation distribution to argue that recent clustering observations 
are consistent with smooth SFHs, and we discuss further tests of our 
model and alternative bursty SFH scenarios.  In Section~\ref{sec:lowz},
we comment on processes that cause SFHs to depart from our 
smoothly-rising, scale-invariant scenario at lower redshift.  
Finally, we summarize our results in Sec~\ref{sec:sum}.

\section{Simulations}
\label{sec:sims}

\subsection{Numerical Simulations} \label{ssec:numsim}
\begin{deluxetable}{lllll}
\tabletypesize{\footnotesize}
\tablecaption{Simulation parameters.}
\tablecolumns{5}
\tablewidth{0pt}
\tablehead{
\colhead{$L$\tablenotemark{a}} &
\colhead{$\epsilon$\tablenotemark{b}} &
\colhead{$m_{\rm SPH}$\tablenotemark{c}} &
\colhead{$m_{\rm dark}$\tablenotemark{c}} &
\colhead{$M_{\rm *,min}$\tablenotemark{c,d}} 
}
\startdata 
$12$ & $0.469$ & $0.235$ & $1.19$ & $15.0$ \\
$24$ & $0.938$  & $1.87$  & $9.55$ & $120$ \\
$48$ & $1.88$   & $15.0$  & $76.4$ & $962$ \\
\enddata
\tablenotetext{a}{Box length of cubic volume, in comoving $\hmpc$.}
\tablenotetext{b}{Equivalent Plummer gravitational softening length, in
comoving $\hkpc$.}
\tablenotetext{c}{All masses quoted in units of $10^6M_\odot$.}
\tablenotetext{d}{Minimum resolved galaxy stellar mass.}
\label{table:sims}
\end{deluxetable} 

We ran our cosmological hydrodynamic simulations using
our custom version of the parallel cosmological galaxy formation
code Gadget-2~\citep{spr02,spr05}.  This code uses an entropy-conservative
formulation of smoothed particle hydrodynamics (SPH) along with a
tree-particle-mesh algorithm for handling gravity.  It accounts for
photoionization heating starting at $z=9$ via a spatially uniform
photoionizing background~\citep{haa01}.  Gas particles undergo
radiative cooling under the assumption of ionization equilibrium,
where we account for metal-line cooling using the collisional
ionization equilibrium tables of \citet{sut93}.  

The assumption
of ionization equilibrium with a uniform ionizing background may
become increasingly inappropriate as we push our predictions 
beyond $z=6$, where cosmological reionization is believed to have 
ended~\citep[for example,][]{fan06}.  We argued in DFO06 that the 
observable galaxies at $z=6$--10 live in sufficiently biased regions 
that their environments were probably reionized early relative to 
the cosmological mean, with the implication that the~\citet{haa01} 
ionizing background is not a strong approximation.  In future work, 
we will relax this assumption through self-consistent radiative 
hydrodynamic simulations.  However, we retain it in our present work
for simplicity, in essence adding this to the list of approximations 
that our comparisons test.  

We endow dense gas particles with a subgrid 
two-phase interstellar medium consisting of hot gas that condenses via 
a thermal instability into cold star-forming clouds, which are in turn 
evaporated back into the hot phase by supernovae~\citep{mck77}.  The 
model requires only one physical parameter, the star formation 
timescale, which we tune in such a way that it reproduces 
the~\citet{ken98a} relation~\citep{spr03}.  

We have improved our treatment of the formation and transport of metals.
We summarise these changes here; for details we refer the reader
to~\citet{opp08}. Star-forming gas particles self-enrich owing to 
Type II supernovae as before, but we now incorporate the 
metallicity-dependent Type II supernova yields from~\citet{chi04} 
assuming a~\citet{cha03} IMF.  We account for mass loss from AGB stars 
using the delayed feedback tables of~\citet{bru03}.  We account for 
energy and metal feedback from Type Ia supernovae (both prompt and 
delayed populations).  Finally, we track the enrichment rates of C, O, 
Si, and Fe separately rather than tracking only the total metal mass
fraction.  Gas particles stochastically spawn star particles via a 
Monte Carlo algorithm in such a way that the growth of stellar mass in 
star particles reflects the underlying star formation rate in the gas 
particles.  Each star particle inherits the metallicity of its parent 
gas particle.  

Galaxy formation models must invoke galactic-scale outflows in order to
avoid the overcooling problem~\citep{whi91}.  We have previously shown 
that models coupling the outflow speed and mass-loading factor (that is, 
the mass of material expelled per unit mass of stars formed) to the 
halo mass are uniquely successful in reproducing a wide variety of 
observations of galaxies~\citep{dav06,fin08,opp10}, and the 
IGM~\citep{opp06, opp08, opp09a,opp09b}.  In this work, we again use 
momentum-driven outflows with a normalization $\sigma_0=150\kms$.  In 
order to improve the fidelity with which we implement momentum-driven 
outflow scalings, we now compute mass loading factors using the velocity 
dispersion of each gas particle's host galaxy rather than the local 
gravitational potential.  The velocity dispersion is in turn computed 
from the host galaxy's baryonic mass using~\citet{mo98}.  For further 
details on the physics treatments in the simulations, 
see~\citet{opp06,opp08,opp10}.

Our fiducial simulation 
volume is a cube $24 \hmpc$ long on each side that uses $512^3$ dark 
matter and star particles.  With a mass resolution limit of 128 star 
particles, this implies that our fiducial volume resolves stellar 
populations more massive than $1.20\times10^8\msun$.  This mass limit 
translates into a limiting UV magnitude via the simulated star formation 
histories (SFHs) and the assumed stellar population synthesis model; 
we will show in Section~\ref{sec:lfs} that this resolution limit is 
well-matched to current observational limits.  Throughout this work, we 
will evaluate resolution convergence by comparing with results from 
additional simulations whose volumes span $12$ and $48 \hmpc$ with the 
same number of particles, for an implied stellar mass resolution limit of 
$1.50\times10^7\msun$ at our highest resolution (see Table~\ref{table:sims}).  

We assume a cosmology where $\Omega_M=0.28$, $\Omega_\Lambda=0.72$, 
$H_0=70\kmsmpc$, $\sigma_8=0.82$, and $\Omega_b=0.046$.  Note that
we do not correct observed photometry to our assumed cosmology because 
the implied corrections are typically small ($\approx0.05$ mag)
compared to the uncertainty inherent in photometric redshifts 
($\approx0.2$ mag).  We measure simulated UV luminosities using a
narrow boxcar filter at 1350 \AA~that is not contaminated by 
Lyman-$\alpha$ emission or absorption by the intergalactic medium
along the line of sight, and we do not $k$-correct observed UV 
luminosities to 1350 \AA~because these corrections are also small.

\subsection{Identifying Simulated Galaxies} \label{ssec:bc03}
We identify galaxies within our simulations as gravitationally-bound
lumps of star and gas particles using {\sc SKID} (see~\citealt{ker05} 
for a description).  We sum the SFRs over each galaxy's gas particles to 
obtain its instantaneous SFR.  We obtain its star formation history by 
examining the ages of its star particles.  We compute its stellar 
metallicity by performing a mass-weighted average over its star 
particles.  We define its gas metallicity as the SFR-weighted mean 
metallicity over its gas particles in order to mimic the metallicities 
that would be measured from its emission lines if this were possible.  
In detail, our use of the mass-weighted stellar metallicity 
may not be completely appropriate for comparison with high-redshift 
observations given that the observed luminosities are dominated 
by the youngest stars.  However, we will justify this choice in 
Figure~\ref{fig:mzr} by showing that the SFR-weighted gas 
metallicity, which dictates the metallicity of the most luminous
stars in each galaxy, typically exceeds the mass-weighted stellar 
metallicity by only 0.2 dex. 

Note that we use {\sc SKID} to identify the galaxies at each redshift 
snapshot independently.  This means that our analysis does not depend 
on our ability to identify each galaxy's progenitors and descendants
(except for the illustrative evolutionary trends in 
Figures~\ref{fig:sfrmstar} and~\ref{fig:mzr}).  Consequently, we refer 
the ``SFH" at a given epoch to the stellar age distribution, independent 
of whether those stars formed in-situ or in smaller progenitors that 
were accreted at earlier times.  While this approach prevents us from 
studying our simulated merger histories,~\citet{guo08} have shown that 
mergers are expected to be subdominant in modulating the SFHs of 
high-redshift galaxies (see also~\citealt{cat10}).  The generally smooth 
SFHs predicted by our model support this view.

We obtain each simulated galaxy's stellar continuum by convolving its 
stellar population with the~\citet{bru03} stellar population synthesis 
models assuming a~\citet{cha03} initial mass function (IMF) and the 
'Padova 1994' stellar evolution models (see~\citealt{bru03} for 
references), interpolating within the tables to 
the correct metallicity and age for each star particle.  We account
for dust using the foreground screen model of~\citet{cal00}, where
each galaxy's assumed reddening $E(B-V)$ is tied to the mean stellar 
metallicity using a local calibration~\citep{fin06}.  We treat the 
effect of absorption by the intergalactic medium (IGM) along the line 
of sight to each galaxy using~\citet{mad95}.  Finally, we convolve 
the resulting synthetic spectra with the appropriate photometric 
filter response curves to produce synthetic photometry.  We have
explicity checked that our photometry is unaffected when we use the
updated population synthesis models of~\citet{cha07} rather 
than~\citet{bru03}.  This is expected 
because the observed Spitzer/IRAC [3.6] flux is not sensitive to the 
contribution of thermally pulsating asymptotic giant branch stars at 
redshifts greater than 6~\citep[see also][]{sta09,lab10}.
Note that, throughout this work, we use $[3.6]$ and $[4.5]$ to denote 
Spitzer/IRAC Channels 1 and 2, respectively, and we use $Y_{105}$, 
$J_{125}$, and $H_{160}$ to denote the WFC3/IR F105W, F125W, and 
F160 bands, respectively.

In an improvement over our previous work, we now add nebular line 
emission to our simulated spectra before accounting for dust and 
the IGM opacity following~\citet{sch09}.  We first compute the 
ionizing continuum luminosity directly from the stellar continuum, 
which automatically accounts correctly for the self-consistently
predicted stellar population 
age and metallicity.  We then compute the luminosity in all hydrogen 
recombination transitions from an upper state with $n\leq50$
using~\citet{sto95}.  This step involves several assumptions.  
First, we use Case B recombination theory and omit Lyman-$\alpha$.  
Second, we assume an electron density $n_e=100$ cm$^{-3}$ and 
temperature $T_e=10^4$K.  This is representative because the 
luminosities vary by $\leq30\%$ in the range $n_e=10^2$--$10^4$ 
cm$^{-3}$ and $T_e=$1000--30,000K.  Finally, we follow~\citet{sch10} 
in assuming that all ionizing photons are absorbed by hydrogen rather 
than dust.  We then add the dominant allowed and forbidden metal 
lines using~\citet{and03}, interpolating linearly to each simulated 
galaxy's metallicity.  Incorporating emission lines in this way 
allows us to make physically-motivated predictions regarding the 
impact of emission lines on observable galaxies at $z\geq6$.  

The most uncertain aspect of our treatment for nebular emission involves
the choice of ionizing escape fraction $\fesc$.  Optimally, one would 
like to constrain $\fesc$ observationally, but current data do not 
permit this~\citep{ono10}.  We adopt the assumption that $\fesc=0$ 
because it yields an upper bound to the effect that nebular emission 
can have on galaxy spectra.  This is equivalent to assuming that 
current $z\geq6$ samples did not bring about cosmological reionization.  
However, we will also argue that current samples probe only the 
brightest $1/3$ of all star formation at $z\geq6$ anyway, hence this 
assumption is not inconsistent with the view that star formation 
dominated the reionization photon budget.  

\section{Rising Star Formation Histories} \label{sec:meanSFH}
\begin{figure}
\centerline{
\setlength{\epsfxsize}{0.5\textwidth}
\centerline{\epsfbox{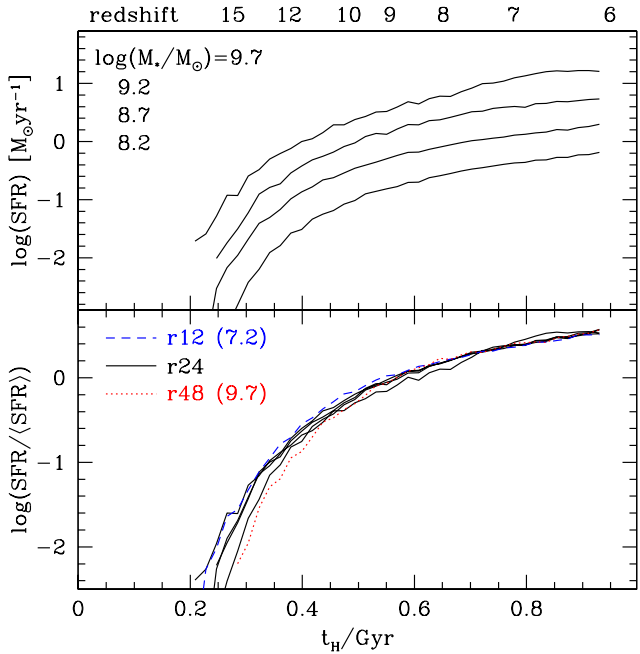}}
}
\caption{Simulated mean SFHs down to $z=5.5$ in four bins of stellar 
mass. \emph{Top:} Mean star formation rates in galaxies whose stellar
masses $\log(M_*/M_\odot)$ at $z=5.5$ are, from bottom to top, 8.2, 
8.7, 9.2, and 9.7.  \emph{Bottom:} The same SFHs but normalized to
the respective mean SFRs.  Our simulations predict that SFHs at early
times are smoothly rising and vary only by a scale factor.  Here we
include representative normalized mean SFHs from our comparison 
volumes from galaxies with the indicated stellar masses.  The good
agreement with the SFHs from our r24 volume suggests that our mean 
SFH is numerically converged.}
\label{fig:meanSFH}
\end{figure}

The goal of the current work is to test two fundamental predictions 
of our galaxy evolution model:
\begin{enumerate}
\item SFHs at early times are smoothly-rising~\citep{fin07}; and
\item SFH shapes are scale-invariant.
\end{enumerate}
We illustrate these predictions in Figure~\ref{fig:meanSFH}.  The top 
panel shows the mean simulated SFHs in four bins of stellar mass.  We 
have generated these SFHs directly from the stellar age distributions 
of our simulated galaxies at $z=6$.  These curves immediately
demonstrate that SFHs are on average smoothly-rising prior to $z=6$. 
The growth timescale $t_0$ in an exponential fit of the form 
$\sfr \propto e^{t/t_0}$ evolves rapidly from 50 Myr at $z=12$ to 350
Myr by $z=6$.

The top panel strongly suggests that mean SFHs of simulated galaxies 
of different masses vary only by a scale factor.  We demonstrate 
this second prediction explicitly in the bottom panel.  Here, we 
have normalized each SFH by its mean SFR, which is just the total 
stellar mass formed by that SFH at $z=6$ divided by the age of the
universe at $z=6$.  The normalized mean simulated SFHs are nearly 
identical, especially after $z=12$.  This implies that galaxy ages 
(and hence colours) should vary only weakly with luminosity, and that 
rest-frame optical and UV luminosities (or, equivalently, the 
inferred stellar mass and star formation rate) should correlate 
with a slope of unity.

This self-similarity may be surprising for two reasons.  First, 
star formation is expected to be quenched at low masses owing to 
photoionization feedback and at high masses owing to the onset of 
hot-mode accretion, hence self-similarity cannot extend to arbitrarily
low and high masses.  Our simulations are sensitive to these effects 
because they include an optically-thin ionizing background and model 
the formation of virial shocks in halos.  However, photoheating only 
suppresses star formation in halos less massive than 
$\sim10^8\msun$~\citep{tho96,oka08} while virial shocks only suppress 
inflows in halos more massive than 
$10^{12}\msun$~\citep{ker05,dek06,ker09}.  The host halos of observed 
galaxies at $z\geq6$ fall largely between these scales 
(Figure~\ref{fig:hod}), hence they do not probe the processes that 
cause departures from self-similarity.

Second, it is surprising that galaxy age should be independent of mass
because more massive dark matter halos are expected to have assembled 
more recently at any redshift~\citep{lac93}.  This might lead one to 
expect more massive galaxies to be younger.  However, \citet{nei06} 
have shown that more massive dark matter halos are younger than 
less-massive halos only if each halo's age refers to the epoch at 
which its most massive progenitor assembles half of the current mass.  
By contrast, if one identifies the halo age with the epoch at which 
more than half of the current mass has been assembled into all 
progenitors more massive than a given threshold, then the trend 
reverses and more massive halos are older.  The stellar age 
distribution relates more closely to the latter age definition if star 
formation is inevitable in halos more massive than, for example, 
$10^8\msun$, hence one might equally expect galaxies in more massive 
halos to be older.  However, the assembled mass fraction in all 
progenitors at a given redshift depends more weakly on mass 
than the mass fraction in the main progenitor (Figures 3 and 4 
of~\citealt{nei06}).  This means that the actual dependence of SFH 
on mass is likely to be driven by feedback processes such as outflows
rather than reflecting underlying trends in halo assembly histories.  
For example, the momentum-driven wind scenario ties the outflow 
mass-loading factor to the circular velocity, which in turn varies 
as $H(z)^{1/3}$~\citep{mo98}; this could delay early star formation 
in massive halos enough to remove the downsizing trend in halo 
formation histories.

Numerical star formation histories suffer from resolution limitations at
early times when the number of gas particles in the protogalaxies is 
small.  Close inspection of the bottom panel of Figure~\ref{fig:meanSFH}
reveals that the normalized SFHs exhibit larger scatter at $z>12$, 
which probably reflects resolution effects.  For this reason, it 
is important to test how sensitive our predicted mean SFH is to the 
simulation's mass resolution. The blue dashed and red dotted curves in 
the bottom panel illustrate normalized SFHs drawn from our higher- and 
lower-resolution comparison volumes, respectively.  They are in 
reasonable agreement with the mean SFHs predicted by our fiducial 
volume.  This suggests that our predicted mean SFH is not strongly 
sensitive to numerical resolution effects.  Moreover, given that the
SFHs appear converged at $z<12$ and that most of the stars observed
during $z=6$--8 formed after $z=12$, the early lack of convergence
does not significantly affect our results.  For reference, the mean SFHs 
in the bottom panel are well-fit by the following polynomial (where $t$ 
is the age of the Universe in Gyr):
\begin{equation}
\sfr \propto \left\{\begin{array}{ll}
  0 & \mbox{if $t<0.27$} \\
  0.2 - 3.2t +10.5t^2-3.9t^3 & \mbox{otherwise}
  \end{array} \right.
\end{equation}

\section{Luminosity Functions}
\label{sec:lfs}

\begin{figure}
\centerline{
\setlength{\epsfxsize}{0.5\textwidth}
\centerline{\epsfbox{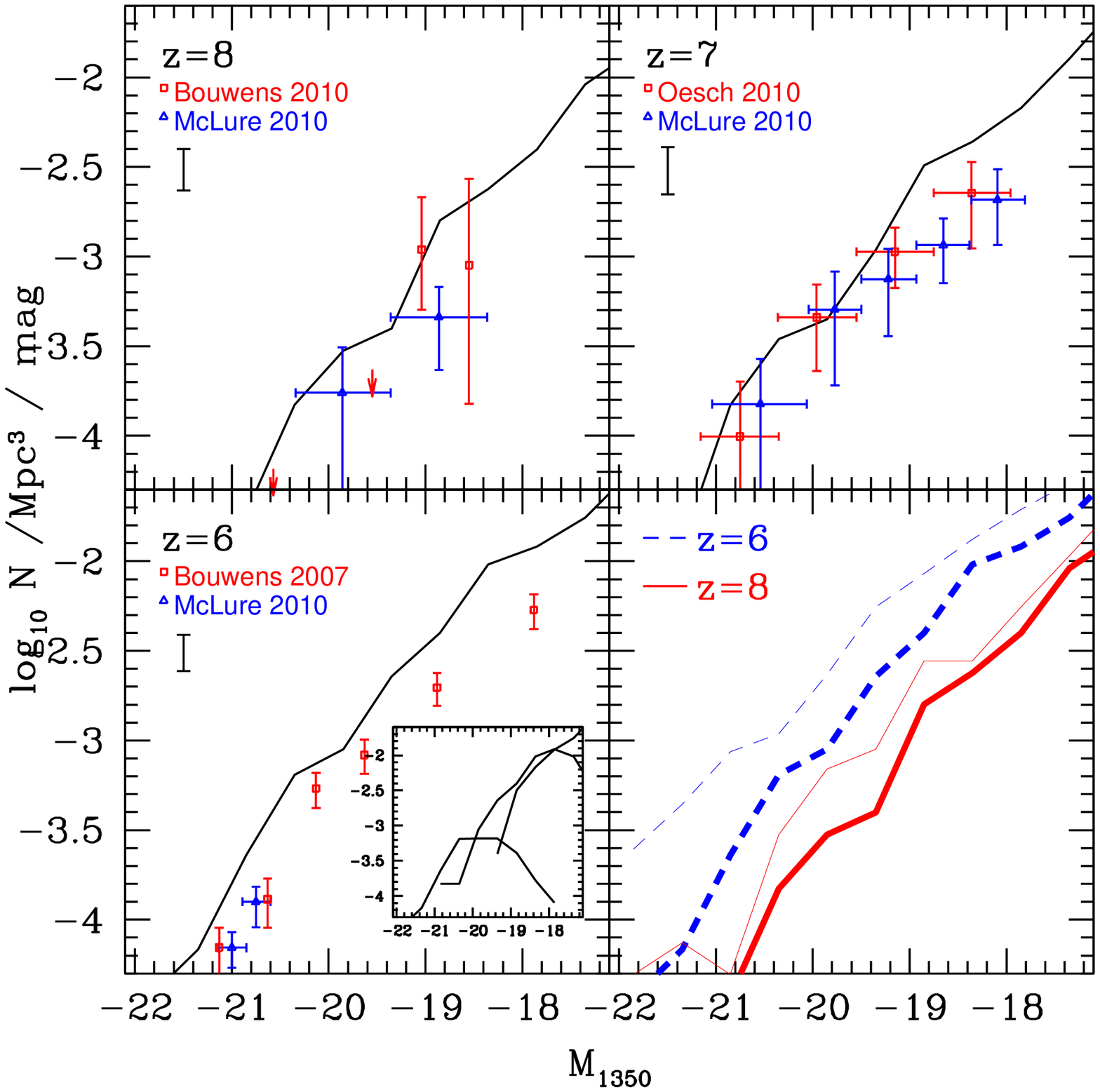}}
}
\caption{Observed (coloured points) and simulated (black) rest-frame UV 
LFs at z=6--8.  The observed LFs are taken from \citet{mcl10} and 
\citet{bou09b} at $z=8$; \citet{oes10a} and \citet{mcl10} at $z=7$; 
and \citet{bou07} and \citet{mcl09} at $z=6$.  The aggregate LFs are 
compiled from our three simulation volumes following DFO06; the inset 
panel in the bottom-left panel shows the three full LFs.  Error bars 
show the jackknife error at $M_{\mathrm{UV}}=-19$ at each epoch; this 
includes Poisson uncertainty and cosmic variance.  Predictions include dust 
extinction following~\citet{fin06}.  The simulations and observations agree 
to within a factor of 3 at all redshifts, suggesting that our star formation 
and feedback prescriptions adequately model galaxy evolution in the late 
reionization epoch.  The lower heavy and upper light curves in the 
bottom-right panel show the simulated LF with and without dust reddening, 
respectively.  The impact of dust is weak owing to low metallicities, 
and it grows progressively weaker with increasing redshift.
}
\label{fig:lfz9876}
\end{figure}

In the top-left, top-right, and bottom-left panels of Figure~\ref{fig:lfz9876}, 
we compare the simulated rest-frame 1350 \AA~LF with the observed UV LFs 
at redshifts 6--8.  Before we discuss these comparisons, we note that 
our predicted LFs are uncertain owing to Poisson errors, cosmic variance, 
numerical resolution effects, and the choice of IMF.  We estimate the 
contribution of Poisson errors and cosmic variance using jackknife 
resampling and show the resulting uncertainties at a representative 
absolute magnitude of $M_{1350}=-19$ in each panel; this is typically 
20--30\%.  We estimate the impact of numerical effects by comparing
the predicted LFs from our three volumes at $z=6$ in the inset panel; 
lower-resolution volumes typically overproduce the LF normalization by 
$\sim50\%$.  Finally, our choice of the \citet{cha03} IMF yields 
luminosities $50\%$ brighter for the same stellar mass than we would 
find from the \citet{sal55} IMF. 

The normalizations of our predicted LFs agree with observations to 
within a factor of three at all three redshifts, which is generally 
within the combined theoretical and observational uncertainties.  In 
particular, both the simulated and the predicted LFs may be regarded 
as brightening by 0.6--0.8 magnitudes during $z=8\rightarrow6$.  
This is a nontrivial test of the smoothly-rising SFH scenario.  
To see this, note that, if the average star-forming galaxy at $z=6$ had 
been forming stars at a smoothly constant or declining rate since it 
formed, then the LF would be constant or grow dimmer to lower redshifts.  
Instead, observations strongly suggest that objects at constant number 
density grow brighter with time, implying that they increase their SFRs 
even as they increase their stellar masses~\citep{pap10}.  This is the 
smoothly-rising SFH scenario. The predicted evolution probes the shape 
of the mean SFH, and the good agreement with observations suggests 
that the simulated SFH shape is realistic.  Note that this 
interpretation depends on the assumption that SFHs are smooth rather 
than, for example, stochastically brightening into and then fading 
out of observed samples~\citep{lee09,sta09}; we will argue on 
observational grounds that the SFHs are indeed smooth in 
Section~\ref{sec:occ}.

While the normalization and its evolution are in reasonable agreement 
with observations, our simulations predict rather steep faint-end slopes 
$\alpha \approx -2.0$.  This is similar to the slope of the halo mass
function, hence our model predicts that low-mass halos possess a fairly 
constant star formation efficiency (defined as the ratio of the SFR to 
the gas mass).  Observations indicate a somewhat shallower slope of
$\approx -1.7$ at $z=6$~\citep[for example,][]{bou07}, which could 
suggest that outflow strengths (or other feedback processes) scale more 
steeply with mass in reality than in our simulations.  By contrast, 
observations at $z=$7--8 indicate a steeper faint-end slope of 
$\approx -2$~\citep{bou10b}, in better agreement with our predictions.  
If this evolution is confirmed, it could indicate that the feedback 
process that flattens the faint end of the LF at $z\leq6$ is overwhelmed 
by the high gas inflow rate at earlier times, as expected in a scenario
where the star formation and outflow rates do not ``catch up" to the
inflow rate until later times~\citep{bou09,pap10}.  

\begin{figure}
\centerline{
\setlength{\epsfxsize}{0.5\textwidth}
\centerline{\epsfbox{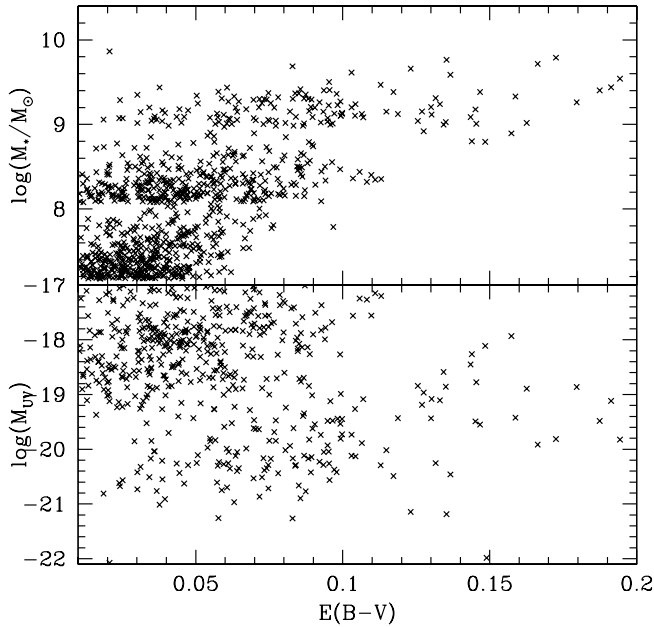}}
}
\caption{The predicted relationship at $z=7$ between dust reddening and 
stellar mass (top) and absolute UV luminosity including the dust (bottom).  
The three loci in each panel reflect our three simulation volumes.  
Typical colour excesses of 0.05 are predicted at all luminosities, 
although a fraction of galaxies more luminous than $M_{UV}<-19$ are 
predicted to have $E(B-V)=$0.1--0.2.
}
\label{fig:dustz7}
\end{figure}

These luminosity functions include the effects of dust, which we add
in post-processing via a prescription that ties the (assumed) dust to 
the (predicted) metallicity following a local calibration~\citep{fin06}.  
We illustrate in Figure~\ref{fig:dustz7} how the resulting colour excess 
$E(B-V)$ varies with stellar mass and absolute UV luminosity (including 
dust) at $z=7$.  Broadly, we predict that $E(B-V)=$0--0.1 for the 
majority of objects at all luminosities although a substantial fraction 
of objects brighter than $M_{UV}=-19$ may have $E(B-V)=$0.1--0.2.  The 
tendency for brighter objects to be dustier reflects the (predicted) 
luminosity-metallicity and (assumed) metallicity-dust relations.

The bottom-right panel of Figure~\ref{fig:lfz9876} illustrates 
how this dust impacts the predicted UV LFs.  Thick and thin 
curves show the predicted UV LFs with and without dust, 
respectively, at $z=6$ (blue dashed) and $z=8$ (solid red).  
Our dust model predicts that dust reddening is weak at early 
times owing to galaxies' low metallicities and the strong predicted 
mass-metallicity relation.  Additionally, the dust column at a given
luminosity is predicted to decrease only modestly with increasing 
redshift because the mass-metallicity and stellar mass-star formation 
rate relations evolve weakly.  For example, at a fixed magnitude of 
$M_{1350}=-19$, $E(B-V)$ increases from 0.04 to 0.06 during this 
epoch.  We will return to the impact of dust reddening in 
Section~\ref{sec:colors}.

\begin{figure}
\centerline{
\setlength{\epsfxsize}{0.5\textwidth}
\centerline{\epsfbox{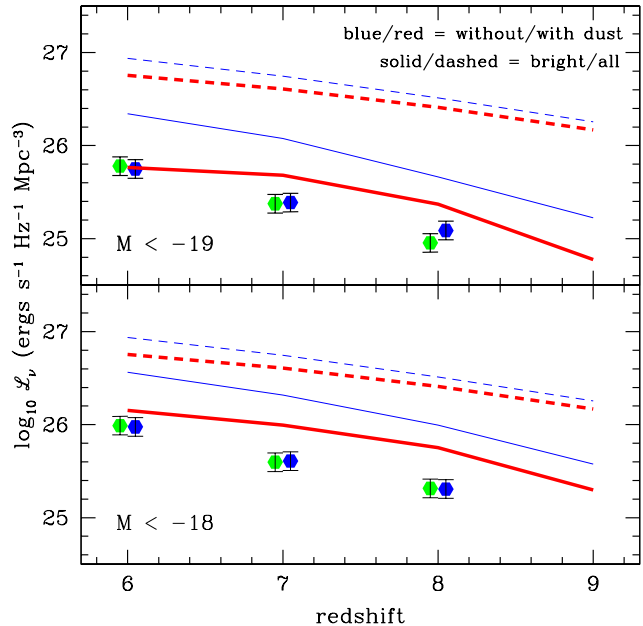}}
}
\caption{Observed (coloured points) and simulated (curves; from the r24 
volume) rest-frame UV luminosity density at z=6--8.  We obtain the 
observed luminosity densities by integrating published luminosity 
functions down to the indicated limits.  The observed luminosity 
functions are from \citet{mcl10} (blue) and \citet{bou09b} (green) at 
$z=8$; from \citet{ouc09} (green) and \citet{mcl10} (green) at $z=7$; 
and from \citet{bou07} (green) and \citet{mcl09} (blue) at $z=6$.  
The two observed densities at each redshift are offset for clarity.  
Solid and dashed curves indicate the simulated luminosity density 
including bright galaxies (down to the indicated limits) and all 
simulated galaxies, respectively, while heavy red and light blue 
curves include and omit dust reddening, respectively.  The 
simulations and observations are in reasonable agreement at all 
redshifts although the simulations slightly overproduce the luminosity 
density at low luminosites and high redshifts, which we ascribe to
 resolution effects.  Broadly, we predict that observations capture 
less than 20\% of the luminosity density at $z\geq6$.
}
\label{fig:ldz9876}
\end{figure}

A key test of galaxy formation models involves the evolution of
the star formation rate density.  This can be inferred from 
the UV luminosity density ($\mathcal{L}_\nu$), which is simply the 
integral of the UV LF.  In Figure~\ref{fig:ldz9876}, we show the 
observed (coloured points) and predicted (dark solid curve) rest-frame 
$\mathcal{L}_\nu$ down to $M_{UV}=-19$ (top) and $M_{UV}=-18$ (bottom) 
at $z=$6--8; see the caption for details.  The simulation is in 
reasonable agreement with observations to $M_{UV}<=-19$ at $z=6$ while 
it slightly overproduces $\mathcal{L}_\nu$ at higher redshifts or when 
integrating to fainter limits.  In both cases, the discrepancies 
were anticipated by Figure~\ref{fig:lfz9876}, where it is clear that 
the simulation preferentially overproduces faint ($M_{UV}>-19$) 
galaxies.  As in Figure~\ref{fig:lfz9876}, the observed discrepancies
are comparable to the uncertainty that is expected from Poisson errors, 
cosmic variance, the IMF, and numerical effects.

The question of whether galaxies can contribute enough ionizing
photons to reionize the Universe depends sensitively on how much
star formation occurs in objects that are fainter than the 
current observational limit of 
$M_{1350}<=-18$~\citep[for example,][]{yan04,yan06}.  We now estimate this 
contribution in two ways.  First, we directly measure the total
predicted $\mathcal{L}_\nu$ as a function of redshift within our 
r24 volume (still including the effects of dust). The dark, lower 
dashed curve in each panel indicates this total.  This curve 
suggests that current observations do not detect more than 
$\leq25$\% of the total star formation at $z\geq6$.  In practice,
this estimate suffers from numerical resolution uncertainty
because the resolution limit of our r24 volume corresponds to a 
luminosity only slightly fainter than $M_{1350}<=-18$.  However,
the r12 volume indicates that the true LF continues smoothly to
much lower luminosities (Figure~\ref{fig:lfz9876}).  Hence we may 
obtain a second estimate for the total luminosity density as follows: 
If we assume that the luminosity density per unit halo mass varies 
slowly with halo mass, then we may compute the unresolved fraction 
by simply computing the ratio of collapsed matter in halos down to 
the typical host halo mass at $M_{1350}=-18$ ($5\times10^{10}\msun$
at $z=6$ in our simulations) versus the amount of collapsed matter 
in halos more massive than the photoheating feedback scale of 
$3\times10^8\msun$~\citep{oka08}.  Using a standard Press-Schechter
calculation, this ratio is 0.079.  However, because our outflow 
model predicts that outflow strength should scale as $M_h^{-(1/3)}$, 
the specific luminosity in low-mass halos is lower, and the 
unresolved fraction should also be smaller.  Correcting for this, 
we estimate that observations down to $M_{1350}=-18$ probe the 
brightest 37\% of the total luminosity density, in reasonable
agreement with our direct numerical estimate.

The lighter solid and dashed curves show how our results change if
we neglect dust extinction.  Dust extinction suppresses the bright 
luminosity density by a factor of 2--4 and the total luminosity
density by $\sim50\%$, where the bright galaxies are more extincted
because they are more metal-rich and hence more dusty by assumption.  
The widening gap between the dusty and dust-free luminosity densities
from $z=7\rightarrow6$ tracks a moderate growth in the normalization
of the mass-metallicity relation.

We now pause to present our predicted conversion between UV 
luminosity and star formation rate.  For SFHs that are smooth on
timescales longer than 10 Myr, this conversion is linear; that is,
$\mathrm{SFR} = c_{\mathrm{SFR}} * L_{\mathrm{UV}}$~\citep{mad98,ken98b}.  
Calculating $c_{\mathrm{SFR}}$ requires knowledge of the galaxy's SFH,
metallicity, and dust extinction.  Within our simulations, the SFH and 
metallicity are predicted self-consistently, hence we may predict
$c_{\mathrm{SFR}}$ in the absence of dust.  Considering only galaxies 
with $M_{UV} \leq -18$ and using a narrow boxcar at 1350 \AA~to compute
the UV luminosity, we find that $c_{\mathrm{SFR}}$ is independent of 
luminosity and equal to $2.0\times10^{28}$ ergs s$^{-1}$ Hz$^{-1}$ 
($\msun$ yr$^{-1}$)$^{-1}$ with a $1\sigma$ scatter of 30\% throughout 
the interval $z=9\rightarrow6$.  This factor assumes the~\citet{cha03} 
IMF for consistency with our simulations.  It may be compared directly 
with conversions that are used to derive star formation rates from 
observed UV luminosities at high redshift.

\begin{figure}
\centerline{
\setlength{\epsfxsize}{0.5\textwidth}
\centerline{\epsfbox{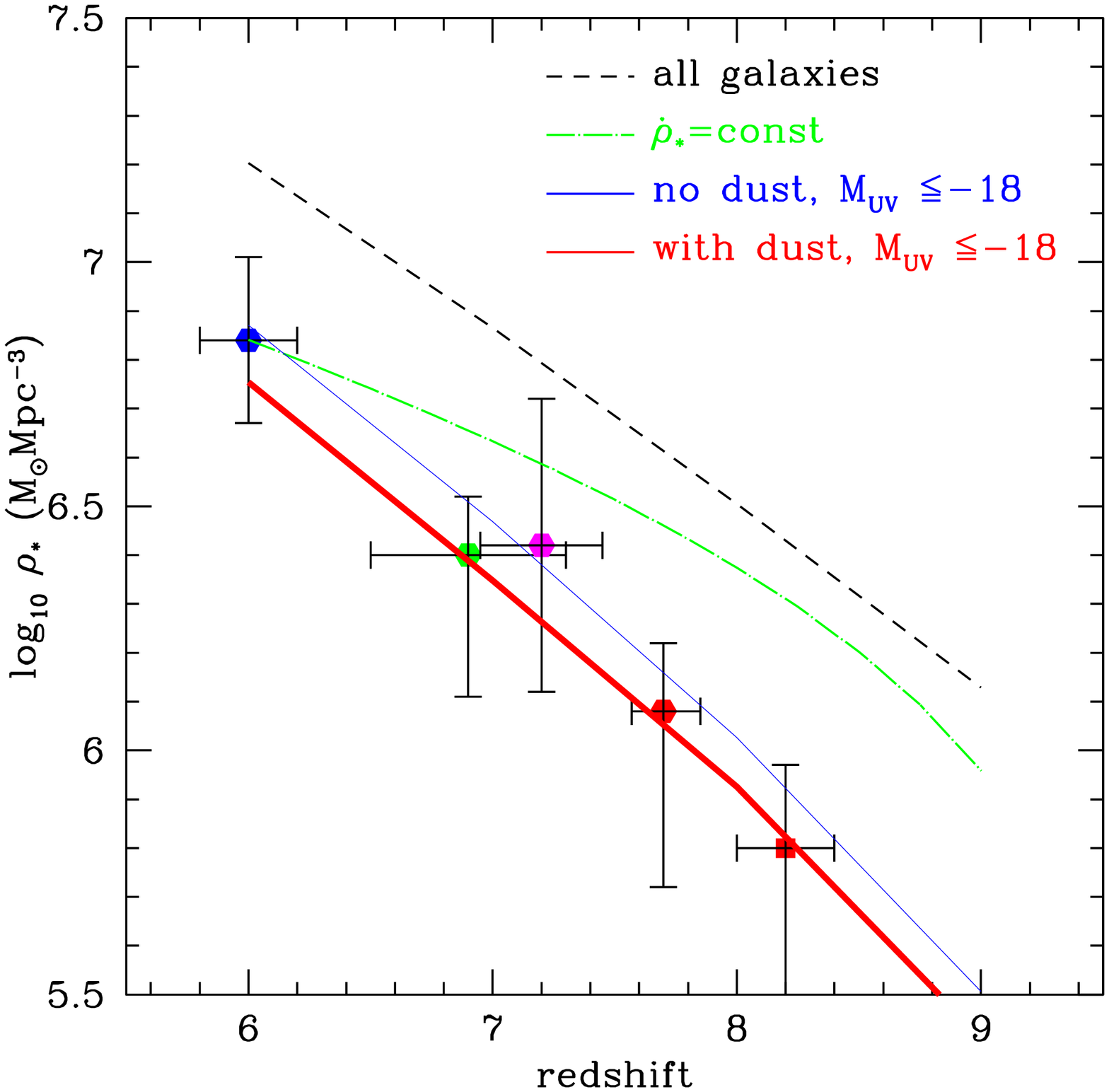}}
}
\caption{Observed (coloured points) and simulated rest-frame
stellar mass density $\rhostar$ in our r24 volume at $z=$6--8.  
The blue hexagon is from~\citet{sta09}; the green hexagon is 
from~\citet{gon09}; the magenta hexagon is from~\citet{lab09}; 
and the red hexagon and square are from~\citet{lab09} 
and~\citep{bou09b}.  Observational points have been corrected to 
account for galaxies brighter than $M_{\mathrm{UV}} \leq -18$ 
following~\citet{lab09}, and we have additionally subtracted 0.18 
dex to convert to a~\citet{cha03} IMF.  Thick red and thin blue 
solid curves show the simulated $\rhostar$ to $M_{UV}<=-18$
including and neglecting dust, respectively, while the dashed 
curve shows the full simulated $\rhostar$.  The green dot-dashed
curve extrapolates the observed $\rhostar$ at $z=6$ backwards
assuming constant SFHs and the simulated star formation rate
density at $z=6$.  The simulated normalization and evolution are
in good agreement with observations.  Moreover, the predicted 
evolution is in much better agreement than non-rising SFH 
scenarios.  Current observations probe 20--40\% of the total 
$\rhostar$ at $z\geq6$.
}
\label{fig:rhostarz9876}
\end{figure}

Returning to our predicted star formation histories,
Figure~\ref{fig:rhostarz9876} compares the simulated and observed stellar 
mass densities $\rhostar$ at $z\geq6$, again integrating only over 
galaxies down to $M_{UV}\leq-18$.  Comparing the heavy solid red curve 
with the coloured points reveals that the simulation is in surprisingly 
good agreement with observations at these epochs.  The agreement in 
the normalizations may be fortuitous.  For example, our r12 volume
predicts a specific star formation rate 30\% lower than the r24 volume
in the range where they overlap, which implies that increasing the
mass resolution of our r24 volume by a factor of eight would boost
the predicted $\rhostar$ by roughly this factor.  However, the 
agreement in the evolution of $\rhostar$ is nontrivial, and supports
the smoothly-rising SFH scenario.  To demonstrate this, we use a green
dot-dashed curve to illustrate how $\rhostar$ would evolve from
$z=6\rightarrow9$ under the hypothesis of constant SFHs.  In this trivial
case, $\dot{\rho}_* (M_{UV} \leq -18)$ is constant and equal to the value 
at $z=6$.  Adopting the observed $\rhostar$ and simulated $\dot{\rho}_*$
for galaxies with $M_{UV}\leq-18$ (including dust) at $z=6$, we find that
the predicted evolution is significantly shallower than observed.  The
evolution is shallower because, in the realistic case of rising SFHs,
galaxies become fainter and drop out of the sample at higher redshifts.
This demonstration is ``conservative" in two senses: First, our simulations
may overproduce $\dot{\rho}_*$ at $z=6$ (Figure~\ref{fig:lfz9876}); repeating
the exercise with a lower $\dot{\rho}_*$ would yield shallower evolution.  
Second, adopting exponentially-decaying SFHs would lead to increasing 
$\rhostar$ with increasing $z$ (because galaxies near the limit 
$M_{UV}=-18$ at one epoch would fade below it at later epochs), in 
serious conflict with observations.  Hence the excellent agreement 
between the predicted and observed slopes is a nontrivial success of 
the smoothly-rising SFH scenario.

As in the case of the UV luminosity density, we estimate observational
incompleteness using our direct numerical predictions.  The dashed 
curve indicates the complete stellar mass density over all galaxies 
in the simulation.  Comparing the heavy solid red and black dashed 
curves suggests that current observations do not detect more than 
20--40\% of the true $\rhostar$ at $z\geq6$.  This is similar to our 
estimate of the detected fraction of the UV luminosity density, as 
expected given that $M_* \propto \sfr$.

\section{Galaxy Colors}
\label{sec:colors}

\subsection{Mean SEDs} \label{ssec:seds}
\begin{figure}
\centerline{
\setlength{\epsfxsize}{0.5\textwidth}
\centerline{\epsfbox{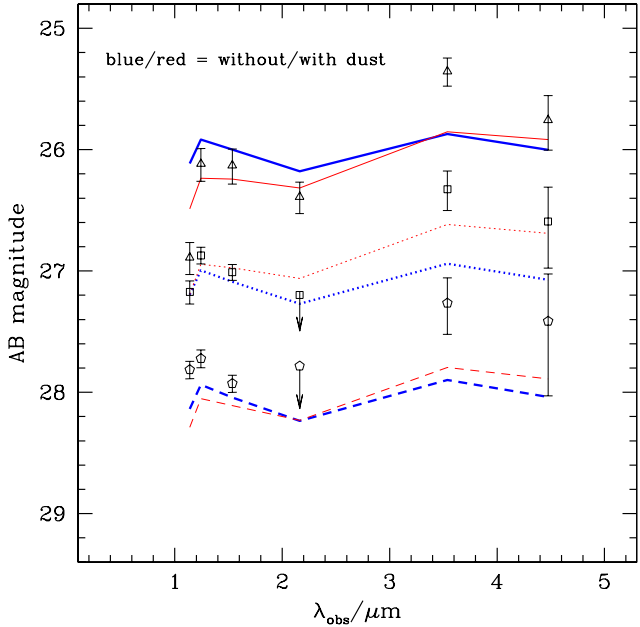}}
}
\caption{Predicted versus observed mean SEDs at $z=7$ in three
bins of $\Delta H = 1$ centered on $H_{160}=26,27,28$.
Simulated SEDs are plotted with solid, dotted, and dashed curves.  
Heavy blue and light red curves correspond to simulations without 
and with dust extinction, respectively.  Triangles, squares, and 
pentagons indicate observations from UV-selected galaxies at 
$z\sim7$~\citep{lab09}.  Note that the set of model galaxies that 
falls within each bin depends on whether dust is present.  This
gives rise to an apparent tendency for the medium and faint bins 
to be brighter in the dusty models longward of $H_{160}$ owing to
their redder SEDs.}
\label{fig:meansedz7}
\end{figure}

In this Section, we study the mean spectral energy distributions
(SEDs) of our simulated galaxies in order to gain some intuition
into their intrinsic properties and their implied selection 
function.  We begin by showing in Figure~\ref{fig:meansedz7}
the mean SEDs of our simulated galaxies at $z=7$ in three bins 
of $H_{160}$ (see caption for details).  Examining the simulated 
intrinsic SEDs first (heavy blue curves), we see that both the
simulations and the observations indicate SEDs that vary weakly
with $H_{160}$, as expected in a scenario where the shape of a
galaxy's SFH varies weakly with its luminosity.  All three bins 
show blue intrinsic UV continua, in good agreement with 
observations~\citep{bou09a,bou09b,bou10a,lab09,fink09}.  By 
contrast, the simulated $[3.6\mu]$ fluxes are noticeably weaker 
than observed at all luminosities.  Adding dust extinction 
reddens the simulated mean SEDs modestly without changing 
the level of agreement with observations.

\begin{figure}
\centerline{
\setlength{\epsfxsize}{0.5\textwidth}
\centerline{\epsfbox{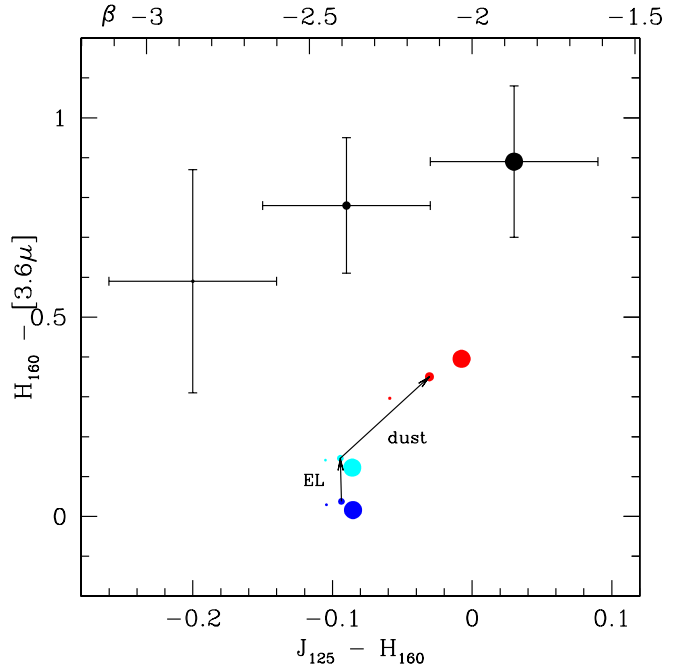}}
}
\caption{Observed (points with errors) versus simulated (points without
errors) colours of galaxies at $z=7$.  Circles indicate the mean colours 
in the same three bins of $H_{160}$ as in Figures~\ref{fig:meansedz7}; 
circle size scales linearly with $H_{160}$ flux.  The effects 
of emission lines (EL) and dust on the predicted mean colours are shown.  
The top axis converts $J_{125}-H_{160}$ to UV continuum slope $\beta$
following~\citet{bou10a}.  The predicted $J_{125} - H_{160}$ colours 
are quite blue as observed although the faintest bin may be too red,
implying that its predicted metallicity is too high.  The simulated 
$H_{160} - [3.6]$ is significantly bluer than observed.  Nebular line 
emission and dust both increase the apparent Balmer break strength, 
but do not close the gap with observations.
}
\label{fig:colcol}
\end{figure}

In order to consider the UV continua and apparent Balmer breaks in
more detail, we compare in Figure~\ref{fig:colcol} the simulated 
$J_{125}-H_{160}$ and $H_{160}-[3.6]$ colours.  This Figure may be
compared with Figure 1 of~\citet{lab09}.  Once dust is included, 
the simulated UV continua (that is, the $J_{125}-H_{160}$ colours) 
in the medium and bright bins are consisent with observations 
owing largely to their young ages (100--150 Myr), low metallicities 
($Z/Z_\odot < 0.5$; Figure~\ref{fig:mzr}), and low predicted dust 
columns.

By contrast, the predicted UV continua in the faintest bin are too red 
even if dust is omitted.  This suggests that their predicted ages or 
metallicities $Z$ are too high.  It is not likely that this 
discrepancy can be resolved purely through younger ages.  This is
because the UV continua of Population II stars are too red even in 
the absence of nebular continuum emission for ages greater than 10 
Myr~\citep[][Figure 3]{bou10a}, and observing populations with ages
younger than 10 Myr is unlikely given that these objects have 
dynamical times of $\sim 100$ Myr.  We note that our predicted ages 
of $\approx 150$ Myr at $z=7$ agree with the recent simulation 
of~\citet{sal10}, despite incorporating different treatments for both 
outflows and metal enrichment.  A more likely possibility is that $Z$ 
suffers extra suppression in low-mass galaxies owing to a steeper scaling 
between the outflow mass loading factor $\eta_W$ and the galaxy mass.  
For example, our simulations currently assume that $\eta_W$ varies 
inversely with the velocity dispersion $\sigma$.  Adopting a dependence 
$\eta_W\propto \sigma^{-\alpha}$ with $\alpha>1$ would further suppress 
$Z$ for faint galaxies because $Z\propto 1./(1+\eta_W)$~\citep{fin08}.  
Such a model would also bring the faint end of the UV LF into better 
agreement with observations (Figure~\ref{fig:lfz9876}).  A final
possibility, as pointed out by~\citet{bou10a}, is a small contribution 
from low-metallicity stars with a top-heavy IMF.  Such a population 
would bluen the UV continuum as long as it did not give rise to 
significant nebular continuum emission, which in turn would require a 
high ionizing escape fraction.  It would be interesting to explore this 
possibility within the self-consistent Population III treatment 
of~\citet{sal10}; unfortunately, these authors did not consider their
predicted photometric colours.

Our models qualitatively reproduce the observed tendency for more 
UV-luminous galaxies to display redder UV continua~\citep{bou09a,bou10a}
owing entirely to the simulated luminosity-metallicity and (assumed) 
metallicity-reddening relations.  Hence the reported colour-magnitude 
trend could imply a modest trend of increasing dust reddening at 
higher UV luminosity, as has previously been observed at lower 
redshifts~\citep{meu99,sha01}.  Note, however, that the presence of 
such a trend in the $z=7$ data remains unclear at present~\citep{sch10}.

Turning to the $H_{160} - [3.6]$ colours, we find that the simulated 
galaxies generically exhibit significant Balmer breaks at all 
luminosities, in qualitative agreement with observations.  We 
previously predicted that significant Balmer breaks are expected in 
Lyman break samples at $z=4$~\citep{fin06}, and Figure~\ref{fig:colcol} 
now extends this prediction out into the reionization epoch.  

In detail, however, the observed breaks are roughly 0.5 mag stronger 
than predicted.  In combination with the blue UV 
continua, these colours pose a challenge for our simulations.  This 
result is similar to the finding by~\citet{lab09} that linearly rising 
SFHs do not produce Balmer breaks stronger than $H_{160}-[3.6]=0.3$ mag
whereas the observations require break strengths of 0.5 mag or stronger.  
One possible explanation is nebular line emission.  For example, 
\citet{sch09,sch10} have used stellar population synthesis models 
including treatments for nebular continuum and line emission to show 
that the strong apparent Balmer breaks observed at $z\geq6$ can be 
mimicked by nebular line emission from extremely young ($<10$ Myr), 
low-metallicity populations~\citep[see also][]{zac08,ono10}.  Our 
simulations provide physically-motivated priors for the typical ages 
and metallicities at these epochs, allowing us to predict the strength 
of the nebular emission lines.  We find that line emission improves 
the agreement with the observed $H_{160} - [3.6]$ colour by $0.1$ mag.  
We do not find that line emission can completely explain the observed 
$H_{160} - [3.6]$ colours because our relatively evolved, enriched 
stellar populations do not yield sufficiently strong emission line 
equivalent widths (in contrast to the wide range of models explored 
by~\citealt{sch09}).  Note that this conclusion is conservative in the 
sense that, in modeling emission lines, we have assumed an ionizing 
escape fraction of zero.  For sufficiently large ionizing escape 
fractions that observed galaxies could contribute significantly to 
reionization (10--50\%), the lines would be correspondingly weaker.  

Another possible explanation for the observed colours is inherently
bursty SFHs.  For example,~\citet{lab09} show that only highly bursty 
SFHs can simultaneously reproduce the observed $J_{125}-H_{160}$ and 
$H_{160}-[3.6]$ colours at low masses (and then only barely).  
Encouragingly, they also find that such a model would not overproduce 
the scatter in the observed SFR-$\mstar$ relation.  Hence 
it is possible that boosting our mass resolution would resolve more minor 
perturbations and instabilities, giving rise to more bursty SFHs and 
redder predicted $H_{160}-[3.6]$ colours.  We can explore this 
possibility by appealing to the tendency for $H_{160}-[3.6]$ to vary 
weakly with luminosity (Figure~\ref{fig:meansedz7}) and examining the 
trend at lower masses in our higher-resolution volume.  We find that 
increasing our mass resolution by a factor of eight boosts $H_{160}-[3.6]$ 
by 0.02 mag on average.  This is quite small compared to the 0.5 mag 
discrepancy with observations, hence if resolution limitations are to 
blame then overcoming them may require ``zoom-in" simulations of 
overdense regions given that our current simulations already probe the 
limits of what is computationally feasible.

It is also possible that the discrepancy owes to a patchy dust scenario in 
which a fraction of the lines of sight through an LBG's interstellar medium 
(ISM) are optically thick to UV; this would suppress the rest-frame UV much 
more strongly than the optical.  However, it is difficult to reconcile this 
scenario with the blue observed UV continua, which seem to imply that these
galaxies are essentially dust-free.

Is it possible that a different treatment for galactic outflows could
suppress the simulated SSFR, thereby deepening the predicted Balmer 
breaks? As can be seen from Figure 2 of~\citet{dav08}, the predicted SSFR 
is relatively insensitive to the details of the outflow treatment.  This 
is because changing the outflow strength broadly increases or decreases 
$\mstar$ and SFR together, without substantially changing their 
ratio~\citep[see also][]{dut09,bou09}.  Hence it is unlikely that the 
predicted $H_{160}-[3.6]$ colours could be made redder by modifying our 
outflow treatment.

Finally, it cannot be ruled out that the photometric uncertainties 
reported by~\citet{lab09} underestimate the true errors because the 
stacked fluxes could be biased by the presence of bright outliers.  For 
example, if (despite their careful analysis) the rest-frame optical 
flux of a fraction of the sources in each bin were contaminated by
incompletely-subtracted neighboring objects, the resulting error 
in the stacked SED would be difficult to detect given the small number 
of sources in each bin.  This reinforces the need for larger samples 
both in order to reduce errors and in order to allow for more accurate 
uncertainty estimates.
 
In summary, our simulations produce blue UV continua and significant
Balmer breaks at all luminosities, in qualitative agreement with
observations.  The faintest observed galaxies are 0.1 mag bluer than 
predicted, which may suggest that our feedback model insufficiently 
suppresses the metallicities of low-mass galaxies.  Our predicted UV 
continua qualitatively reproduce the reported colour-magnitude trend
as long as we include our metallicity-dependent dust reddening 
prescription.  The simulations produce apparent Balmer breaks that 
are weaker than observed, even if we correct for resolution effects
and account for nebular line emission.  This is the most dramatic 
discrepancy that we have found, and it emphasizes the need for a 
better observational understanding of the nature of the observed 
apparent Balmer breaks at $z\geq6$ through $K$-band imaging, 
mid-infrared spectroscopy, or mid-infrared imaging with improved 
spatial resolution.

\subsection{Direct Spectral Energy Distribution Fitting} \label{ssec:spoc}
\begin{figure}
\centerline{
\setlength{\epsfxsize}{0.5\textwidth}
\centerline{\epsfbox{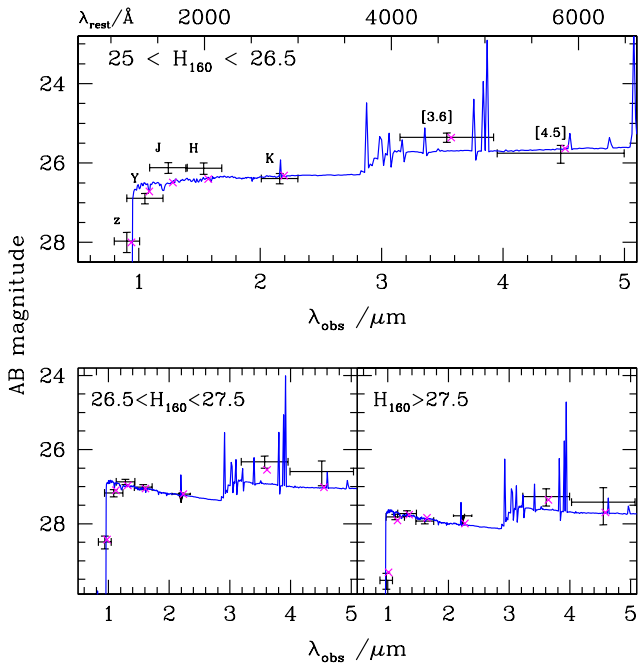}}
}
\caption{Black points with errors denote the stacked SEDs of $z\sim7$ 
LBGs from~\citet{lab09}; the $H_{160}$ magnitude ranges for each bin
are indicated.  The blue curve indicates the spectrum of the best-fit 
numerically-simulated galaxy.  Magenta crosses indicate synthetic 
photometry of that spectrum, offset slightly in wavelength for clarity.  
The simulated spectra yield reasonable agreement at all wavelengths,
although the WFC3/IR measurements in the brightest bin are not well-fit.
Optical emission lines readily explain the observed $[3.6]-[4.5]$ 
colours.
}
\label{fig:sedfit}
\end{figure}

In Section~\ref{ssec:seds}, we compared the predicted and observed
colours of galaxies at $z\sim7$ without allowing the redshift or the 
amount of dust extinction to vary.  This approach is justified
if the photometric redshifts are accurate and the dust extinction is 
negligible.  On the other hand, given that we do not model the dust 
extinction self-consistently and that the redshift is not 
well-constrained, a complementary approach is to allow these 
parameters to vary freely and ask how well the simulations can account 
for observations in principle.  In this Section, we use our Bayesian 
SED-fitter {\sc spoc}~\citep{fin07} to determine how well the 
smoothly-rising SFHs fit observed SEDs at $z\sim7$ and discuss the 
implied physical properties.  

Here we briefly review {\sc spoc}; for details and tests 
see~\citet{fin07}.  {\sc spoc} operates similarly to other SED-fitters.  
Starting with a set of model SFHs, it generates a library of SEDs by 
resampling the SFHs over a grid of redshift $z$ and dust column $A_V$.  
Note that, in an improvement over our previous work, we include optical 
emission lines
as described in Section~\ref{ssec:bc03}.  After comparing the library 
SEDs with the measurements, it uses a Bayesian analysis to derive the 
posterior probability for the various physical parameters from the 
likelihoods ($\exp(-\chi^2/2)$) and any priors.  The novelty of this 
approach is that we directly adopt numerically simulated SFHs and 
metallicities rather than assuming toy-model SFHs (for example, 
constant or declining).  This imposes physically-motivated priors 
on the results because the prior probability of a given combination 
of age, metallicity, $\mstar$, and SFR is proportional to the number 
of model galaxies that form with these parameters.  It is for this 
reason that, in contrast to conventional SED-fitting studies, {\sc spoc} 
only resamples the model SFHs in $z$ and $A_V$.  The number of free 
parameters is difficult to define given that the simulations do not 
sample parameter space uniformly.  However, given that metallicity
and SFR correlate tightly with $\mstar$ whereas age varies only with
redshift, there are effectively only three, namely $\mstar$, $z$, and
$A_V$.  The resulting likelihoods are equally as interesting as the 
constraints because they quantify how well numerical SFHs can account 
for the full SEDs of individual objects, automatically identifying 
observed colours that challenge the models.

As a test of how well our numerical simulations can account for the full 
SEDs of reionization-epoch galaxies, we apply {\sc spoc} to the three 
binned SEDs at $z\sim7$ from~\citet{lab09} in order to take advantage 
of their high signal-to-noise.
We derive our model SFHs from snapshots of our 24 $\hmpc$ volume at a 
number of discrete redshifts between $z=6$--8.  We do not consider 
low-redshift solutions~\citep[see][]{mcl09}.  Perturbing each model 
galaxy in redshift and sampling 100 different values of $A_V$ between 
0--1.5, we generate a library of 5 million model SEDs for comparison.
We treat detections with less than $2\sigma$ significance as $2\sigma$ 
upper limits and adopt the reported measurements and errors in the 
other bands.  We adopt the WFC3/IR $Y_{105}$ and VLT/ISAAC $K_s$ bands 
to compute simulated $Y$ and $K$ fluxes, respectively.

In Figure~\ref{fig:sedfit}, we compare the observations (black points) 
with the spectrum of the best-fit combination of model SFH, redshift, 
and $A_V$ (blue curve) and its associated photometry (magenta crosses).  
Examining the brightest bin first (top panel), we find that our 
best-fit SED matches the observed fluxes in all bands other than $J_{125}$ 
and $H_{160}$, both of which it underproduces by $\approx2\sigma$.  
Intriguingly, we find that optical emission lines completely explain 
the observed $K-[3.6]$ and $[3.6]-[4.5]$ colours.  Many reionization-epoch 
SEDs show evidence for a blue optical continuum suggestive of optical 
emission lines~\citep{lab09,gon09}.  Our modeling suggests that these 
colours are real rather than reflecting observational uncertainty (for 
example, in the IRAC measurements), with several implications.  First, 
it reinforces the need for SED-fitting studies to account for nebular 
emission~\citep{sch09,sch10}.  Second, it surprisingly suggests that 
the difficulty with the observed $H_{160}-[3.6]$ colours 
(Figure~\ref{fig:colcol}), if observational in origin, may owe as much 
to the WFC3/IR as to the IRAC bands given that the $K-[3.6]$ and 
$[3.6]-[4.5]$ colours are not difficult to reproduce.  Finally, it 
reinforces the need for deep $K$-band imaging in order to constrain 
the strength of the Balmer Break.

We may derive constraints from our posterior probability distributions
by marginalizing over all but one parameter at a time and determining 
the resulting 68\% confidence intervals.  This method generally suffers 
from significant degeneracies between the derived age, metallicity, and 
dust extinction~\citep[for example,][]{sha01,pap01}.  However, restricting 
our attention to the SFHs that arise within our simulations partially
breaks these degeneracies because the predicted range of SFHs is narrow,
effectively introducing tight priors (for details, see~\citealt{fin07}).  
We find $z=6.7\pm0.1$, $A_V=0.6\pm0.1$, $\log(\mstar/\msun)=9.6\pm0.1$, 
$\sfr=16.0\pm3.8\msun\mbox{yr}^{-1}$, and $Z=0.007\pm0.0009$.  The 
inferred SFR is 
60\% larger than inferred by~\citet{lab09}, and the inferred stellar 
mass is 10\% lower (both numbers are corrected for the different assumed 
IMFs).  Meanwhile, the inferred dust column is much larger than would be 
expected from the blue $J_{125}-H_{160}$ colour.  This explains why the
model SED is too red in $H_{160}-[3.6]$ whereas, with our fiducial dust 
prescription, it was too blue (Figure~\ref{fig:colcol}).  Not surprisingly,
the associated $J_{125}-H_{160}$ colour is now also too red whereas it
was in good agreement with the observed colour in Figure~\ref{fig:colcol}.
These constraints are driven largely by the $K$--$[4.5]$ bands, and the 
lower stellar mass is permitted in part by the inclusion of nebular 
emission lines~\citep[for example,][]{sch10,ono10}. 

In the bottom panels, we compare the stacked SEDs from the medium and 
faint bins of~\citet{lab09} with the best-fit model SEDs.  These bins 
yield generally better fits to our models than the brighter bin.  As 
before, the simulated SEDs readily reproduce the observed $[3.6]-[4.5]$ 
colours owing to the presence of optical emission lines.  The medium 
stack implies the physical parameters $z=6.8\pm0.1$, $A_V=0.08\pm0.08$, 
$\log(\mstar/\msun)=8.8\pm0.1$, SFR=$3.1\pm0.4\msun\mbox{yr}^{-1}$, and 
$Z=0.0040\pm0.0006$, while the faint stack corresponds to $z=6.9\pm0.1$, 
$A_V=0.09\pm0.08$, $\log(\mstar/\msun)=8.5\pm0.1$, 
$\sfr=1.5\pm0.3\msun\mbox{yr}^{-1}$, 
and $Z=0.0032\pm0.0007$.  As before, the stellar masses are lower than 
would be inferred without emission lines.  Meanwhile, the inferred SFRs 
are 10--30\% lower than inferred by~\citet{lab09} (after accounting 
for the different IMFs) owing to the low predicted metallicities.  The 
inferred dust columns are significantly lower than in the bright bin.  
Note, however, that deeper observations in $K$ may eventually demand 
higher dust columns in the faint bins as well.

\subsection{Survey Completeness} \label{subsec:completeness}
We may use our simulated SEDs to predict the completeness of 
current surveys.  In detail, of course, completeness varies 
with the selection strategy.  However, a tight correlation 
between stellar mass and SFR implies that the completeness of
UV-selected samples is dominated by the detection limit rather than 
by colour cuts~\citep{fin06}.  This is especially true when the
dust column is low.  For example, the colour cuts in~\citet{oes10a} 
do not eliminate any of our simulated galaxies at $z=7$.  

\begin{figure}
\centerline{
\setlength{\epsfxsize}{0.5\textwidth}
\centerline{\epsfbox{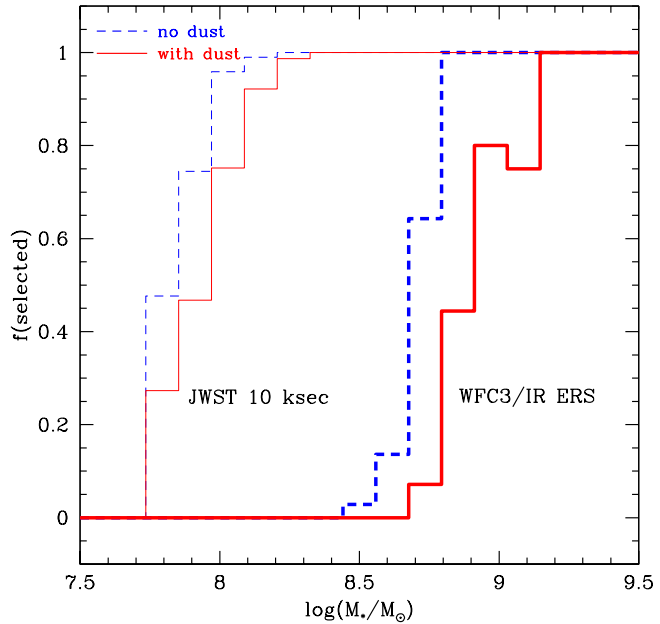}}
}
\caption{Completeness versus stellar mass at $z=7$.  Curves
show how the number fraction of galaxies that are selected by 
two different criteria varies with stellar mass in simulated
catalogs with (solid red) and without (dashed blue) dust 
reddening.  Heavy curves correspond to the current $5\sigma$ 
limiting magnitude in the WFC3/IR ERS fields of
$J\leq27.7 \wedge H\leq27.4$~\citep{bou10a}.  Light curves 
correspond to the $5\sigma$ detection limit for a 10,000 second 
exposure in the JWST/NIRCAM F200W band.
Current surveys detect most galaxies more massive than $10^9\msun$
and few that are below $10^{8.5}\msun$, while JWST will easily
detect objects that are 10\% as massive.
}
\label{fig:completenessz7}
\end{figure}

In Figure~\ref{fig:completenessz7}, we plot using heavy curves
the number fraction of galaxies that are selected by the $5\sigma$ 
detection limits in $J_{125}$ and $H_{160}$ of~\citet{bou10a} versus 
stellar mass both with and without dust.  Galaxies with 
$\mstar\geq10^9\msun$ are selected efficiently assuming that dust 
reddening is small ($E(B-V) \leq 0.05$) owing to their high star 
formation rates.  Dust increases the 50\% completeness limit from 
$10^{8.7}\msun$ to $10^{8.9}\msun$ by extincting galaxies whose 
UV luminosities are already near the detection limits.  Galaxies 
with stellar masses less than $10^{8.5}\msun$ are not selected 
efficiently regardless of dust because their star formation rates 
are too low.  The $5\sigma$ detection limit for a $10^4$ second 
exposure through the JWST/NIRCAM F200W filter will be $K=29.72$;
\footnote{\url{http://www.stsci.edu/jwst/instruments/nircam/sensitivity/index_html}}
we show how this improves the mass completeness using light curves.
Clearly, a relatively shallow JWST exposure will readily detect 
objects that are 10\% as massive as the least massive objects 
detected in the WFC3/IR ERS fields.  The impact of dust extinction 
will be weak owing to the low metallicity (Figure~\ref{fig:mzr}).

These results are robust to our choice of input
physics or mass resolution.  This is because the mass completeness 
is mostly sensitive to the specific star formation rate, which in
turn is relatively insensitive both to our choice of outflow 
strength and mass resolution.

\section{Physical Properties}
\label{sec:phys}

In this section, we summarize the predicted physical properties 
of our simulated galaxies at $z\geq6$, where possible comparing 
with observational inferences.  The most important physical 
process impacting these predictions is our simulation's 
implementation of momentum-drive outflows, and we have previously 
explored the effects at $z=6$ of varying our outflow 
treatment in DFO06.  

\begin{figure}
\centerline{
\setlength{\epsfxsize}{0.5\textwidth}
\centerline{\epsfbox{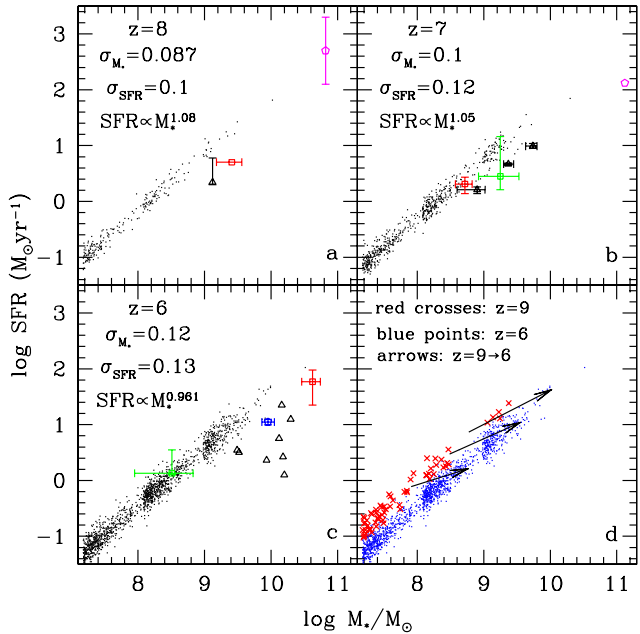}}
}
\caption{Panels a--c show simulated (black points) and observed (coloured 
shapes) SFR-$\mstar$ relationships.  The three simulated loci correspond 
to our three volumes, and their good agreement suggests that numerical 
errors are weak.  We include the slope and dispersion about the best-fit 
linear relation from the r24 volume at each redshift.  Panel d shows the 
simulated relationships at $z=9$ (red crosses) and $z=6$ (blue points) 
as well as typical evolution of individual galaxies from $z=9\rightarrow6$
(arrows).  The red square at $z=8$ is from~\citet{bra08}, where we assume 
constant star formation and 
$Z_*=0.0004$; the black triangle corresponds to the best-fit SFH to 
the $z=7.7$ stack of~\citet{lab09}; and the magenta hexagon is Object 1 
from~\citet{cap09}.  At $z=7$, black triangles are the stacked results 
from~\citet{lab09}; the red square is from~\citet{ega05}, where the 
point represents the average of the inferences assuming a 100 Myr 
exponentially decaying SFH over the different assumed metallicities, 
and the errors reflect this error added in quadrature to assumed 
photometric errors of (5\%, 20\%) in the (SFR, mass); the magenta 
hexagon is Object 3 from~\citet{cap09}; and the green square is the
$z=6.6$ LAE stack of~\citet{ono10}, where we plot result from the
``pure stellar" model with $Z/Z_\odot=0.2$.  At $z=6$, the blue 
pentagon is LA34 from~\citet{lai07}; the red square is 
from~\citet{dow05}; the black triangles are from~\citet{eyl07}; and
the green square is from the stacked SED at $z=5.7$ of~\citet{ono10} 
assuming ``pure stellar" models with $Z/Z_\odot=0.2$.  We have 
subtracted 0.18 dex from all observationally inferred SFR and $\mstar$ 
to adjust from the Salpeter to the Chabrier IMF.  Our simulations 
predict that SFR $\propto \mstar$, with tight scatter and a 
weakly-evolving normalization.  The slope and evolution of the simulated 
trend are in good agreement with observations, but the normalization 
may be offset by a factor of 2--3.
}
\label{fig:sfrmstar}
\end{figure}

\subsection{Star Formation Rates} \label{sec:sfrmstar}
We show in Figure~\ref{fig:sfrmstar} how SFR relates to $\mstar$
in our simulated catalogs (black points) at four different redshifts.  
Broadly, we predict SFR$\propto \mstar^{\mbox{0.95--1.05}}$ at all 
redshifts, confirming our previous results (DFO06).  The predicted
slope at $z=7$ of 1.05 is in excellent agreement with the reported 
slope of $1.06\pm0.1$~\citep{lab09}, and the agreement is good at the
other epochs as well.  This comparison is a nontrivial test of our 
model, and given that our simulations were not tuned to reproduce 
such a trend---in fact, this class of simulations predicted 
it~\citep{dav00,wei02}---the level of agreement is remarkable.  

There are two possible interpretations for a near-unity slope.  One 
possibility is that galaxies begin forming stars at different epochs, 
but once they begin, their growth is exponential with a timescale 
$\mstar/\sfr$.  This timescale is observed to be $\approx 500$ Myr at 
$z=6$~\citep{sta09}, although our simulations indicate somewhat 
shorter timescales of 200--350 Myr (note that accounting for optical 
emission lines will bring observations into closer agreement with our 
predictions; Section~\ref{ssec:spoc}).  The short growth timescales 
could then imply a bursty star formation scenario in which galaxies
``turn on" at different times, form stars with a short duty cycle and
a constant exponential growth timescale, and then fade into a quiescent
state~\citep{sta09}.  There are two difficulties with this picture.
One is that it is not obvious why galaxies that begin forming at very
different times should nonetheless obey the same exponential growth 
timescale with negligible scatter.  The other is that bursty models 
predict a large population of quiescent galaxies at $z\geq4$ that 
has not been observed~\citep{bra07}.  We will return to these points 
in Section~\ref{sec:occ}.

A second interpretation of the near-unity slope is that SFHs at 
$z\geq6$ have a scale-invariant shape.  This is because, if all 
galaxies begin forming stars at the same time and with the same 
SFH shape, then the SFR and the $\mstar$ both vary linearly with 
an overall scale factor.  Consequently, the slope remains near 
unity until a mass-dependent process such as hot-mode accretion 
or AGN feedback breaks the scale-invariance and flattens the 
slope~\citep{dav08}.  In this smoothly-rising SFH scenario, the 
growth timescale is dominated by smooth inflows, which in turn are 
regulated by the competition between the growth of halo potential 
wells and the decrease in the cosmic density~\citep{bou09,dut09}.  
It is not easy to distinguish observationally between the bursty 
and the smoothly-rising scenarios using UV-selected samples, 
although our models (and, for that matter, all hydrodynamical
simulations) support the latter view.  However, we will discuss
other observational probes in Section~\ref{sec:occ}.

The predicted normalization may be slightly offset from observations.  
This offset could have two possible implications.  First, it suggests 
that observationally inferred stellar masses are too high because they 
do not include the effect of optical emission lines.  Accounting for 
this would likely reduce observationally inferred stellar masses by 
0.1--0.3 dex (Figures~\ref{fig:colcol} and~\ref{fig:sedfit}), improving 
the agreement with predictions.  In support of this view,~\citet{lab09}
have found that applying an observationally-motivated estimate for 
the strength of the OII emission line lowers the typical inferred 
$\mstar$ at $z=7$ by 0.17 dex.  The second implication is 
that our simulated SFRs are slightly too high, which we previously 
noticed evidence for in Figure~\ref{fig:lfz9876}.  It is tempting 
to suppose that strengthening our outflows would solve this problem.  
Unfortunately, as noted in Section~\ref{sec:colors}, this would not
suppress the predicted SSFR because increasing the outflow mass-loading
factors reduces $\mstar$ and $\sfr$ together without changing their 
ratio~\citep[see also][]{dut09}.  Hence it is not simple to understand 
how the bulk of the offset could be attributed to our input physics.  

The predicted dispersion about the mean trend is $\approx0.1$ 
dex at all redshifts.  This tight scatter reflects the tendency for 
star formation to be driven by smooth gas accretion rather than by 
interactions~\citep{ker05,bir07}.  It is somewhat less than the reported 
scatter of 0.25--0.3 dex at $z\sim7$~\citep{lab09}.  Given that 
observational uncertainties boost the observed scatter, it is possible 
that the true scatter is consistent with our predictions.  It is also 
possible that our limited numerical resolution does not account 
fully for the minor interactions or discrete infalling clouds that 
perturb galaxies away from their equilibrium SFR.  We evaluate this
possibility by plotting the predicted SFR-$\mstar$ trend from all 
three of our simulation volumes at each redshift (hence the three 
separate ``clumps" in the simulated locus).  These three volumes 
span a factor of 64 in mass resolution and should immediately reveal 
which predictions are strongly sensitive to numerical effects.  The 
offsets in the trends are small compared to the scatter, 
and the scatter does not vary with resolution.  This
indicates that the predicted trends are numerically converged.  
Moreover, it is interesting to note that the mass resolution of our 
$12\hmpc$ simulation is 50\% \emph{higher} than the gas-phase mass 
resolution with which~\citet{mih96} demonstrated that mergers can 
boost SFRs of MW-scale galaxies by factors of 10--100.  The fact 
that our predicted SFR-$\mstar$ scatter remains small despite our 
resolving power supports the view that the predicted scatter is
robust to resolution limitations, and that dramatic merger-driven 
starbursts are indeed uncommon at high redshift.

\citet{dut09} use a semi-analytic model for the growth of disk galaxies 
within a smoothly-accreting halo to predict a scatter of $\sim0.1$ dex 
at all redshifts.  They attribute this to scatter in the 
mass accretion histories of the host halos.  They then speculate that 
the larger observed scatter could owe to scatter in the gas accretion 
rate at fixed halo mass, which their model does not account for.  Our 
simulations account for this effect self-consistently, hence the fact 
that our predicted scatter agrees with theirs argues that the observed 
scatter is not dominated by scatter in the gas accretion rate at fixed
halo mass.

In the bottom-right panel, we show how galaxies' masses and SFRs evolve 
from $z=9\rightarrow6$ in each of our simulation volumes.  Red crosses
and blue points show the simulated loci at $z=9$ and 6, respectively.  
Comparing these loci reveals that the SFR at a given $\mstar$ (that is, 
the normalization of the $\mstar$-$\sfr$ relation) is predicted to 
decline by $\approx0.3$ dex from $z=9\rightarrow6$.  As we have already 
seen, observations also indicate that this normalization evolves 
weakly out to at least $z=8$, qualitatively supporting this 
picture~\citep{gon09,sta09}.  This slow evolution indicates that 
galaxies grow roughly exponentially during this epoch, as can also be 
seen from the shallow evolution of the slope of the mean SFH after 
$z=9$ in Figure~\ref{fig:meanSFH}.  As pointed out by~\citet{sta09}, 
the only smooth SFH that is consistent with these trends is a rising 
one.  

In order to illustrate how this works, we have added arrows indicating 
the typical evolution of individual galaxies during this epoch.  We 
trace this evolution by looping through the 10 most massive galaxies 
identified in each simulation at $z=6$ and searching for their most 
massive progenitors at $z=7$--9.  We then average the masses and SFRs 
of these 10 galaxies at $z=6$ and their most massive progenitors at 
$z=9$; the resulting average evolution is representative of how all 
galaxies are predicted to evolve.  Galaxies evolve in a direction that 
is slightly shallower than the mean trend at a given redshift.  The 
tendency to evolve nearly parallel to the observed trend indicates 
smoothly-rising, nearly exponential growth while the slightly 
shallower slope reflects a 
slow decline in gas accretion rates owing to cosmological 
expansion~\citep{bou09}.

Observational inferences suffer from two systematic biases, either of 
which could introduce artificial offsets or scatter at each redshift.  
First, the relevant data were measured by many different groups (see 
the caption), each of which introduces slightly different photometric 
techniques and selection biases.  The reported errors may be 
underestimated~\citep{sch10} although side-by-side comparisons 
suggest that this problem is not large~\citep{fink09}.  Second, 
different groups used different assumptions in SED-fitting.  This 
introduces scatter because the choice of SFH and stellar metallicity 
biases the inferred stellar mass by 10--30\% on average, with biases 
of up to a factor of 10 possible for certain model 
SFHs~\citep[for example,][]{sha01,pap01,fin07,mar10}.  It would be of interest 
to reproduce these observational constraints using unified modeling 
assumptions at all epochs in order to minimize these 
effects~\citep[for example,][]{sch10}.  Viewed differently, however, 
it is intriguing that, despite the variety of observational and
modeling techniques that underlie these observations, the resulting 
constraints tell a consistent story: The observed SFR-$\mstar$ 
trend obeys a near-unity slope, its normalization evolves slowly, 
and it may be slightly offset to lower SFR or higher $\mstar$ than 
predicted by our model.  It is not unreasonable to suppose that 
applying a consistent set of modeling assumptions to these data 
would only tighten and reinforce the inferred trends.

In summary, current observations indicate that our simulations
reproduce the observed slope and evolution of the SFR-$\mstar$
trend at $z\geq6$, while the normalization may be somewhat offset
and the scatter is 50\% lower than observed.  The near-unity slope 
supports the view that observed reionization-epoch galaxies began
forming stars at similar epochs and possess SFHs that differ 
from one another only by a scale factor.  The offset in the 
normalization, if confirmed, indicates that the inferred stellar 
masses are too high because they do not correct for optical 
emission lines.  The weak observed evolution in the normalization is 
inconsistent with any smooth SFH other than a rising one.  The small 
predicted and observed scatters are broadly consistent with the
view that high-redshift star formation is driven predominantly by 
gas inflows rather than mergers, while the fact that the observed 
scatter is larger may reflect observational uncertainties.  

\subsection{Metallicities} \label{sec:metmass}
Direct observations of gas-phase abundances now indicate that 
galaxies exhibit progressively lower metallicities at higher 
redshifts~\citep{erb06,mai08}.  In DFO06, we showed that this occurs 
naturally in the hierarchical growth scenario and predicted that 
observable galaxies at $z\sim6$ should exhibit metallicities less 
than one tenth solar.  Since then, a number of works have lent 
qualitative support to this prediction by noting that subsolar 
metallicities yield better fits to observed SEDs at $z\geq6$ than 
solar metallicities~\citep[for example,][]{eyl07,sta09,bou10a,lab10}.

On the other hand, an important result from DFO06
was the prediction that very little of the observable star 
formation at $z\sim6$ should result in primordial-metallicity
``Population III" stars because metal enrichment occurs so
quickly.  Given that our updated simulations incorporate 
significantly more realistic treatments for metal enrichment 
(Section~\ref{sec:sims}), it is of interest to determine 
whether these results have changed.  To this end, we use 
this section to update our predicted mass-metallicity relation 
at $z\geq6$ and show that improving our treatments for metal 
enrichment revises our predicted metallicities \emph{up} rather 
than down, supporting our view that reionization-epoch galaxies 
should not exhibit a significant contribution from 
Population III stars.

\begin{figure}
\centerline{
\setlength{\epsfxsize}{0.5\textwidth}
\centerline{\epsfbox{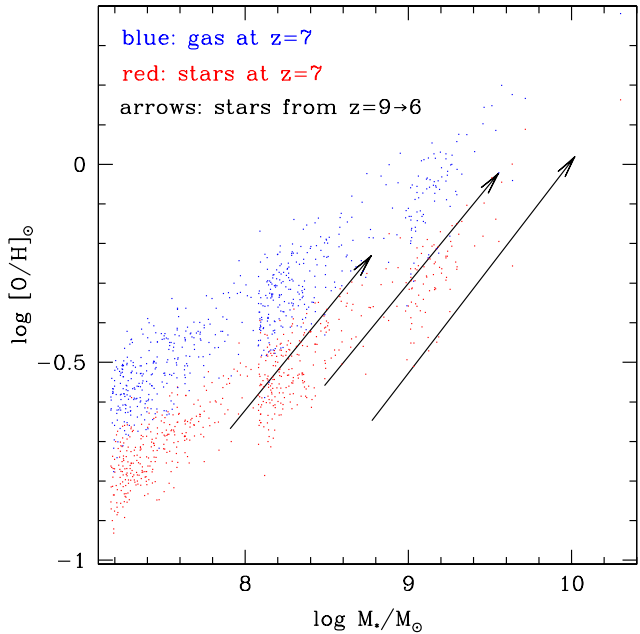}}
}
\caption{Simulated SFR-weighted gas metallicity (blue top locus) and
mass-weighted mean stellar metallicity (red bottom locus) versus stellar 
mass at $z=7$.  We assume a solar photospheric oxygen mass fraction 
of 0.00574~\citep{asp09}.  The three ``clumps" indicate our three 
simulation volumes.  The arrows indicate the mean evolution in stellar
mass and stellar metallicity of the 10 most massive galaxies in each 
simulation volume from $z=9\rightarrow6$.  Observable galaxies are 
already enriched above $0.1Z_\odot$ by $z=7$.  Their evolution is 
slightly steeper than the mean trend at a given epoch owing to weak 
evolution in the normalization of the mass-metallicity relation.
}
\label{fig:mzr}
\end{figure}

In Figure~\ref{fig:mzr}, we show with red points (bottom locus) the 
mass-weighted mean stellar oxygen metallicities of our simulated 
galaxies at $z=7$, normalized to the solar photospheric oxygen mass 
fraction.  Broadly, we predict that observed reionization-epoch 
galaxies possess metallicities greater than $0.1Z_\odot$.  We also 
predict a tight trend with $\approx 0.1$ dex of scatter in the residuals, 
similar to our previous finding at $z=2$~\citep{fin08}.  Note that 
the predicted relation for the iron mass fraction is similar but shifted 
down by 0.2 dex, reflecting the weak contribution of prompt Type Ia 
supernovae at early times.

The upper, blue locus shows the SFR-weighted mean gas-phase oxygen 
metallicities for the same galaxies.  These points predict that 
gas metallicities follow the same trends as the stellar metallicities 
but boosted by $\approx0.2$ dex becase stellar metallicities reflect 
the lower metallicities that characterized each galaxy's progenitors.  
This figure may be compared to Figure 5 of DFO06, where we used a 
similar set of simulations to predict that the gas-phase metallicities 
at $z=$6--8 were generally larger than 1/30$Z_\odot$.  Our current 
simulations track metal enrichment significantly more
realistically, as summarized in Section~\ref{sec:sims}.  Given this 
abundant increase in realism, the fact that our newer simulations 
predict higher metallicities than our previous work underscores the 
conclusion that observable stellar populations at $z=$6--9 should 
not contain a significant stellar mass fraction below the 
Population III threshold of $[O/H]<-3$~\citep{bro04}.

We use arrows to indicate how galaxies evolve in stellar mass and
stellar metallicity from $z=9\rightarrow6$ in each of our three 
simulation volumes (the gas-phase metallicity is similar, but shifted
up by 0.2 dex).  We compute this mean evolution using the 10 most 
massive galaxies in each volume
at $z=6$ as described in Section~\ref{sec:sfrmstar}.  Galaxies evolve 
along a direction that is slightly steeper than the mean trend at $z=7$.  
This can be understood as follows~\citep[see also][]{fin08}: In the 
presence of strong outflows, galaxy metallicities closely track the 
equilibrium metallicity $Z=y/(1+\eta_W)$, where $y$ is the metal yield 
and $\eta_W$ is the mass-loading factor, or the rate at which material 
enters the outflow divided by the SFR.  This equilibrium metallicity 
owes to strong coupling between the inflow, star formation, and outflow 
rates.  In the momentum-driven wind scenario, $\eta_W$ shrinks with 
increasing mass, hence the equilibrium metallicity grows with increasing 
mass.  The tendency for galaxies to evolve slightly more steeply than 
the mean trend at a given epoch reflects the prediction that the 
normalization of the mass-metallicity relation grows by 0.2 dex from 
$z=9\rightarrow6$.

To understand the increase in our predicted metallicities with respect to 
DFO06, note that, at a fiducial stellar mass of  $10^9\msun$ and redshift 
$z=6$, DFO06 predicted a gas-phase metal mass fraction of 0.002,
whereas our current simulations predict a mass fraction of 0.006 overall, 
with 0.005 of this in oxygen alone.  This increase owes to the various
improvements within both our simulations and our analysis: The newer 
code predicts noticeable $\alpha$ enhancements at $z=6$; the adopted 
mass-loading factors were twice as large at a given halo mass in our 
previous simulations; and our adoption of a~\citet{cha03} IMF when 
computing the yields significantly boosts the enrichment rates.  Each 
of these effects boosts the predicted gas-phase metallicities by 
0.1--0.3 dex, hence an overall factor of three increase is not surprising.  
This difference may be regarded as an estimate in the uncertainty owing 
to our imperfect understanding of how metals are created and transported 
(although we believe that our current simulations are the more realistic).  
For example, if we were to re-run our simulations with higher mass 
loading factors in order to improve the agreement between the simulated 
and observed UV LFs (Figure~\ref{fig:lfz9876}), our predicted 
metallicities would shrink by 0.1--0.3 dex.  Broadly, however, this 
uncertainty is far too small to change the fundamental conclusion from 
Figure 5 of DFO06: Observable stellar populations at $z=$6--9 are robustly 
predicted to have metallicities that are well in excess of 0.01$Z_\odot$, 
hence it should be possible to understand their SEDs without reference to 
the predicted SEDs of Population III stars.  

\begin{figure}
\centerline{
\setlength{\epsfxsize}{0.5\textwidth}
\centerline{\epsfbox{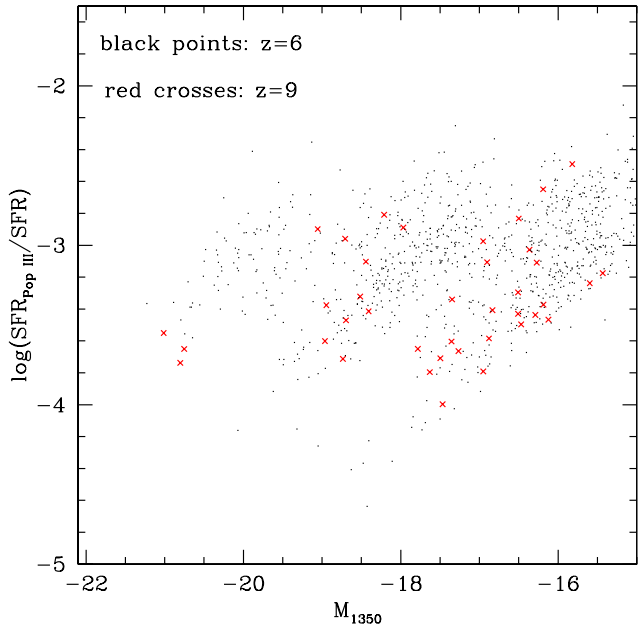}}
}
\caption{The fraction of star formation occurring in gas with oxygen 
mass fraction less than $10^{-3}Z_\odot$ as a function of absolute 
magnitude at $z=9$ (red crosses) and $z=6$ (black points).  A residual 
fraction of Population III star formation is predicted to persist even 
in bright galaxies at $z=6$.
}
\label{fig:PopIII}
\end{figure}

Despite the fact that star formation at $z\geq6$ is dominated by 
metal-enriched gas, a residual fraction of metal-poor star formation is 
predicted to persist even in observable galaxies.  To estimate this, we 
have computed for each simulated galaxy the fraction of star formation 
occurring in gas whose oxygen mass fraction falls below 
$10^{-3}Z_\odot$.  In Figure~\ref{fig:PopIII}, we show how this 
fraction varies with absolute magnitude at $z=6$ (points) and $z=9$
(red crosses).  At both redshifts, we predict that roughly $10^{-3}$ of 
all star formation occurs in metal-poor gas, fairly independent of
luminosity.  Hence it is entirely possible that the observed 
reionization-epoch samples contain a tiny fraction of Population III 
star formation although the impact on the total SEDs is negligible.  
In detail, comparison of the three simulated loci suggests that our 
simulations do not completely resolve the metal-mixing processes that 
determine the Population III fraction.  Extrapolating the trend from 
the highest-resolution simulation, we find that galaxies with 
$M_{UV}=-17$ may have a Population III fraction as high as 1\%.  This 
population may eventually be detectable through its enhanced rate
of pair production SNe or GRBs.  Note that these results 
are in qualitative agreement with previous suggestions that a residual 
level of Population III star formation persists after 
$z=6$~\citep{tor07,joh10,mai10,tren09}.

\citet{sal10} have recently used observed samples at $z=5$--10 to 
test the predictions of a cosmological hydrodynamic simulation that
differs from ours in three major respects: First, they incorporate 
an explicit treatment for the transition from Population III to 
Population II star formation whereas our simulations do not treat 
Population III.  Second, they treat the formation, destruction, 
and dispersal of dust grains in detail in order to predict the dust 
reddening whereas our model ties the normalization of a foreground 
dust screen to local observations.  Finally, their simulations model 
galactic outflows under the assumption that the mass-loading factor 
and wind speeds do not vary~\citep{spr03} while our simulations 
assume momentum-driven winds~\citep{mur05}.  Despite these differences,
their finding that the impact of Population III stars on observable 
$z=6$--8 galaxies is slight ($\leq1\%$ of the UV luminosity for 
$M_{UV} < -18$) is qualitatively consistent with our own results, and 
justifies our decision to neglect the associated processes in our 
simulations.  

\section{Interpreting the Observed Halo Occupancy} \label{sec:occ}
\begin{figure}
\centerline{
\setlength{\epsfxsize}{0.5\textwidth}
\centerline{\epsfbox{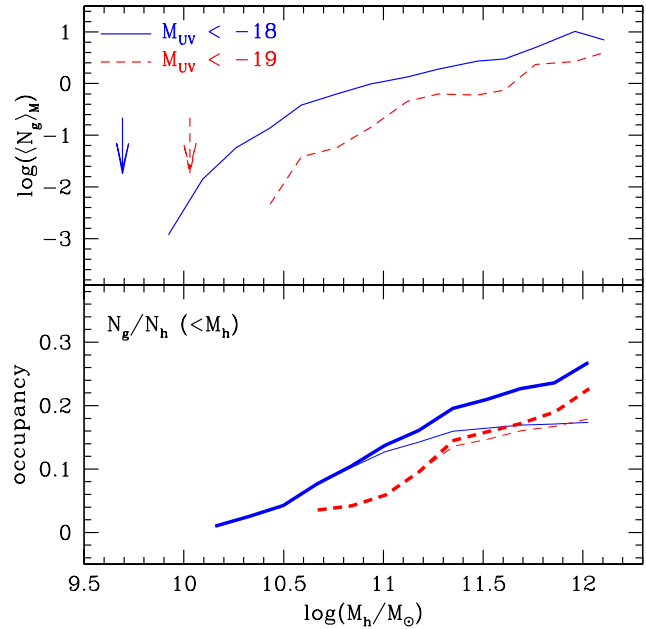}}
}
\caption{\emph{top} The simulated LBG HOD at $z=5.5$ for samples 
brighter than $M_{1350}\leq-18$ (solid blue) and -19 (dashed red).  
The arrows indicate the hypothetical minimum halo mass for each 
sample (see text).  The simulated HOD lies well below unity for 
low-mass halos $\log(M_h/\msun) < 11$ owing to strong outflows, and it 
climbs gradually owing to significant scatter in the relationship 
between halo mass and $M_{1350}$.
\emph{bottom} The cumulative occupancy for samples satisfying the same 
luminosity cuts as in the top panel.  Thick and thin curves show the 
cumulative halo occupancy with and without additional galaxies in halos 
that host more than one galaxy satisfying the luminosity cut.  The 
predicted occupancies lie within the observationally inferred range 
of 0.15--0.6~\citep{lee09}, but they reflect a suppressed HOD owing 
to outflows rather than a star formation duty cycle.
}
\label{fig:hod}
\end{figure}

Up until this point, we have used a variety of arguments to demonstrate
that, if Lyman Break galaxy (LBG) SFHs are smooth, then observations 
indicate that they must be rising rather than constant or declining.  
We have not yet considered the possibility that LBG SFHs are highly 
bursty.  For example,~\citet{lee09} recently used accurate clustering 
observations at $z=4$--6 to show that the fraction of massive halos 
that host LBGs (hereafter, the ``occupancy") is likely less than unity, 
and interpreted this as evidence that LBGs have short 
star formation duty cycles.  Subsequently,~\citet{sta09} built upon 
this idea to speculate that the progenitors of LBGs observed at one 
epoch may not be observable at earlier epochs since the progenitors' 
SSFRs would violate the observed non-evolving SSFR.  This interpretation 
echoes~\citet{fer02}, who used rest-frame UV-optical measurements 
combined with toy-model SFHs to demonstrate that the progenitors of 
LBGs at $z=3$ cannot dominate the observed star formation rate density 
at $z=4$ if SFHs are smooth.  In this Section, we explore how our 
simulations populate dark matter halos with LBGs and argue that outflows 
naturally give rise to the observed occupancy, with the implication 
that there is no conflict between clustering observations and the 
smoothly-rising SFH scenario.  We defer a more direct comparison of our 
predicted clustering properties with observations to future work.

\subsection{Halo Occupation Distribution} \label{ssec:hod}

For reference, we begin by reviewing how an observed sample's 
occupancy may be inferred from its LF and clustering behavior.  The 
connection is made by modeling the underlying galaxy population's 
halo occupation distribution (HOD;~\citealt{ber02,bul02}), or the 
mean number of observable galaxies as a function of halo mass.  This 
in turn derives from the survey selection probability, the LF of 
isolated halos, and the subhalo mass function~\citep{yan03,lee09}.  
To be explicit, let us define the isolated halo LF $dN_g/dL (M)$ as 
the intrinsic number of galaxies with luminosity between $L$ and 
$L+dL$ in a halo of mass $M$; the selection probability for galaxies 
of luminosity $L$ as $p(L)$; and the number of subhalos with mass 
between $m$ and $m+dm$ contained within a halo of mass $M$ as 
$dN_{\mathrm{sh}}/dm (M)$.  Then the mean number of observable 
galaxies as a function of halo mass $\langle N_g \rangle_M$ is
(dropping the $L$ and $M$ dependencies for clarity):
\begin{equation}
\label{eqn:hod}
\langle N_g \rangle_M = 
  \int_0^\infty \frac{dN_g}{dL}p dL +
  \int_0^\infty p dL 
  \int_0^M dm \frac{dN_{\mathrm{sh}}}{dm} \frac{dN_g}{dL}
\end{equation}
The first and second terms in Equation~\ref{eqn:hod} describe the 
contributions of central and satellite galaxies, respectively.
Generally, one assumes that $p(L)$ can be modeled and that the halo
and subhalo mass functions are known.  In this case, HOD modeling 
consists of devising a functional form for the isolated halo LF and 
then constraining its parameters to reproduce the observed luminosity 
function and clustering properties.  For reference, we refer to the
intrinsic LF of individual halos including the subhalo contribution 
as the conditional luminosity function (CLF).

After constraining the HOD, it is straightforward to inquire what 
fraction of the halos that are eligible to host an observable galaxy 
actually do so.  If it is assumed that all halos more massive than 
the lowest-mass halo that contributes to the sample $M_{\mathrm{min}}$ 
are eligible, then this occupancy is given by
\begin{equation}\label{eqn:occupancy}
\frac{N_g}{N_h} = \frac{
  \int_{M_{\mathrm{min}}}^{\infty} \frac{dN_h}{dM} \langle N_g \rangle_M dM}{
  \int_{M_{\mathrm{min}}}^{\infty} \frac{dN_h}{dM} dM},
\end{equation}
where $dN_h/dM$ is the (parent) halo mass function.  Note that,
throughout this discussion, we use ``occupancy" to denote the mean 
number of observable galaxies hosted by halos within a certain mass 
range, including the contribution of their subhalos.  By contrast, the 
``duty cycle" of~\citet{lee09} refers to the integral of the LF over 
luminosity for a single halo, neglecting subhalos.  For sufficiently 
small assumed scatter in the luminosity--halo mass relation, these 
quantities are the same.  

Observationally, the mean occupancy turns out to be robustly less than 
unity at $z=$4--5~\citep{lee09}.  Unfortunately, the physical implication 
of this result is unclear because the nature of the galaxies that inhabit 
``dark" halos is unconstrained.  One possibility is that all halos above 
$M_{\mathrm{min}}$ do, in fact, host a star-forming galaxy and that a 
fraction $1-N_g/N_h$ are in a temporarily quiescent state.  In this case, 
the occupancy $N_g/N_h$ constrains the star formation duty 
cycle~\citep{gia01,lee09,tre10}.  Another possibility is that the 
galaxies in the ``dark" halos are actively star-forming but are too dusty 
to satisfy LBG colour selections.  
In this case, the occupancy constrains the scatter in the dust column
distribution.  To see this, note that randomly rendering half of all 
actively star-forming galaxies too dusty to satisfy LBG selection 
criteria would shift the inferred HOD down by 50\%.  For sufficiently
steep low-mass cutoffs, this would not change $M_{\mathrm{min}}$, 
hence the ratio $N_g/N_h$ would also decrease by 50\%.  It is also 
possible that not all halos above $M_{\mathrm{min}}$ host galaxies whose 
masses are comparable to those of LBGs (LBG ``cousins").  In this case, 
$\langle N_g \rangle_M$ constrains the CLF and a low observed occupancy
has two possible interpretations: Either the CLF has a low normalization 
and many halos do not host galaxies at all, or it has a steep faint end 
and ``dark" halos host low-mass galaxies that are too faint to be 
observed.  We will now demonstrate that our simulations naturally yield
this last scenario, and that there is therefore no conflict between 
a low observed occupancy and a smoothly-rising SFH scenario.

In order to illustrate this possibility, we shall consider our simulated 
LBG HOD at $z=5.5$.  To derive it, we identify dark matter halos using 
the friends-of-friends group finder 
{\sc fof}\footnote{\url{http://www-hpcc.astro.washington.edu/tools/fof.html}}
with a linking length 
tied to the virial overdensity at $z=5.5$.  We then match the halo and 
galaxy catalogs and assign galaxies to parent halos.  Finally, we 
count the number of simulated galaxies brighter than absolute magnitude 
cuts of -18 and -19 that live within each halo.  This simple approach 
does not isolate subhalos from parent halos, hence it only allows us to 
compute the sum of the two terms in Equation~\ref{eqn:hod}.  However, 
this is sufficient for our present purposes; we leave a more detailed 
analysis of our simulated HOD for future work.

In the top panel of Figure~\ref{fig:hod}, we show the resulting 
predicted HODs.  Broadly, the simulated LBGs live in halos 
with masses in the range $\log(M_h/M_\odot)=$10--12.  Brighter 
galaxies live preferentially in more massive halos as observationally
inferred~\citep[for example,][]{gia01}.  Both HODs rise steeply from a
luminosity-dependent minimum halo mass $M_{\mathrm{min}}$, then turn 
over by $\log(M_h/\msun)=11$, then rise as a power law to higher 
masses.  $M_{\mathrm{min}}$, which we define as the mass of the least 
massive halo that hosts at least one galaxy satisfying the luminosity 
threshold, is $\log(M_{\mathrm{min}}/M_\odot)=$10.0 and 10.5 for 
$M_{1350}<-18$ and $-19$, respectively, and the corresponding power 
law slopes are $0.8\pm0.1$ and $1.0\pm0.2$.  

The fact that our HOD lies well below unity at low masses has two 
important implications.  First, our simulated HOD does not resemble 
conventional ``central-satellite" models in which nearly all halos 
more massive than $M_{\mathrm{min}}$ host $\geq1$ 
galaxy~\citep[for example,][]{ber03,kra04,zhe05} because roughly half of the 
simulated LBGs satisfying current detection limits live in halos less 
massive than the mass scale at which the HOD crosses unity.  Second, 
it implies that the mean occupancy (Equation~\ref{eqn:occupancy}) is 
also less than unity.  We confirm this in the bottom panel, which 
gives the cumulative occupancy at each halo mass for all halos between 
$M_{\mathrm{min}}$ and that mass.  Our predicted occupancies of 
0.15--0.3 are consistent with the observationally inferred range at 
$z=4$--5 of 0.15--0.60~\citep{lee09}.  However, they do not correspond 
to a star formation duty cycle because our simulated duty cycle is 
essentially unity 
(Figures~\ref{fig:completenessz7} and~\ref{fig:sfrmstar}).  Instead, 
they indicate that our simulated CLF does not assign an observable 
galaxy to each halo above $M_{\mathrm{min}}$.  This possibility 
reconciles the smoothly-rising SFH scenario with the observed 
clustering behavior of LBGs, but it raises the question of why our 
predicted HODs climb so much more slowly than would be expected in 
conventional central-satellite scenarios.

The answer lies in the impact of outflows.  DFO06 showed that, in the 
absence of outflows, halos at $z=6$ are expected to contain roughly 
their global fraction of baryons.  These baryons invariably collapse 
into galaxies on a dynamical timescale, leading to the conventional 
HOD picture such as the SPH model of~\citet{ber03}.  By contrast, 
momentum-driven winds evacuate an increasing fraction of the baryons 
at lower halo masses because the mass-loading factor $\eta_W$ scales 
inversely with the velocity dispersion.  This mass dependence boosts 
the scatter in the relationship between baryonic mass and halo mass 
because the instantaneous outflow strength depends on both the galaxy 
mass and the redshift, with the result that a galaxy's time-averaged 
outflow strength depends on its entire growth history.  Consequently, 
a significant fraction of isolated central galaxies that would be 
observable in the absence of outflows are suppressed below 
observational limits, rounding off the low-mass end of the HOD and 
suppressing the predicted occupancy.  Broadly, these effects are 
expected to be significant if the outflow mass-loading factor 
$\eta_W > 1$.  For galaxies brighter than $M_{\mathrm{UV}}=-18$, our
simulation assigns mass-loading factors of 3--7, hence this condition 
is satisfied for all observable galaxies at $z=5.5$.  Note that the 
possibility that feedback could dramatically alter the predicted HOD 
was also raised by~\citet{ber03} and~\citet{zhe05}.  

While we have argued that scatter in halo formation histories dominates 
the halo occupancy at low halo masses, two other processes could scatter 
galaxies below $M_{1350}=-18$.  First,~\citet{yos02} have shown that, 
in the absence of outflows, halos consisting of fewer than 75 dark matter 
particles show significant scatter in their gas fractions owing to 
numerical effects.  Such an effect would suppress the predicted occupancy
at low halo masses.  This effect should be negligible in our simulations 
because the smallest halo that hosts an observable galaxy contains 
roughly 1000 dark matter particles, indicating that our resolution cut 
is conservative.  Second, intrinsic scatter in the relationship between 
dust column and halo mass has a similar effect because it selectively 
suppresses the luminosities of the central galaxies of some low-mass 
halos below detection thresholds.  Our dust prescription does contain 
intrinsic scatter, and the HODs in Figure~\ref{fig:hod} include this 
effect.  Neglecting dust extinction boosts the predicted occupancies 
for the fainter sample from 0.3 up to 0.4.  Hence the predominant 
factor suppressing the predicted occupancy of low-mass halos is 
outflows rather than numerical effects or dust extinction.

\subsection{Conditional Luminosity Function} \label{ssec:clf}

\begin{figure}
\centerline{
\setlength{\epsfxsize}{0.5\textwidth}
\centerline{\epsfbox{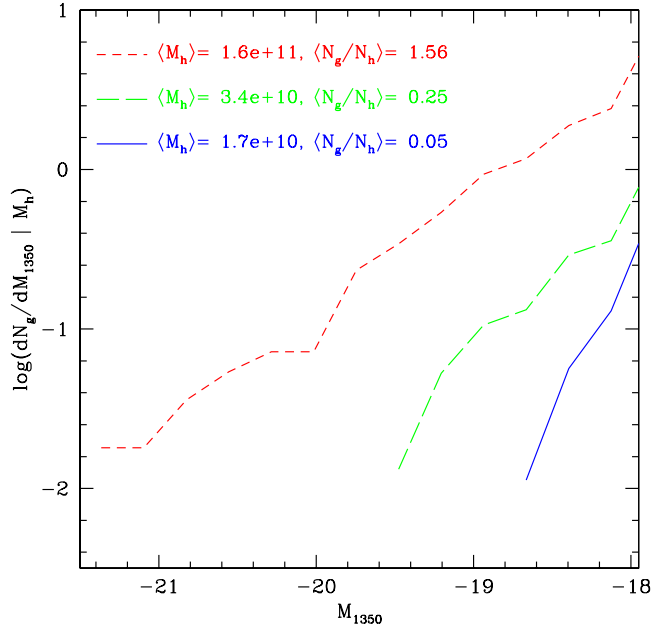}}
}
\caption{The simulated CLF at $z=5.5$ in three bins of halo mass;
mean halo masses and occupancies within these bins are indicated.
The CLF is well-approximated by a power law whose slope steepens and 
normalization decreases to lower masses.  The mean occupancy increases 
with increasing halo mass as observed, but in our simulations this 
reflects the mass-dependence of the CLF rather than a star formation 
duty cycle.
}
\label{fig:clf}
\end{figure}

If ``dark" halos do not host LBGs that are experiencing a temporarily 
quiescent phase, then what sort of galaxies do they host?  We propose 
that halos that are more massive than $M_{\mathrm{min}}$ but that do not 
host an observable galaxy preferentially host galaxies that are simply 
faint.  In Figure~\ref{fig:clf}, we illustrate this possibility through 
the predicted CLF at $z=5.5$ in three bins of halo mass.  The halo mass
bins are $1.26<M_h<2.5$, $2.5<M_h<5$, and $5<M_h$ in units of 
$10^{10}\msun$; the corresponding mean halo masses are indicated in the
Figure.  For each bin, we compute the predicted CLF by counting the 
number of galaxies as a function of absolute UV magnitude (including 
dust) whose host halo mass falls within that mass bin and then dividing 
by the number of simulated halos in that bin.  Our CLF is 
well-approximated by a power law whose slope flattens and 
normalization increases with increasing halo mass.  Unlike the wind-free 
models of~\citet{zhe05}, it does not show a bump at high luminosities 
corresponding to the central galaxy contribution because the scatter in 
the relation between halo mass and UV luminosity smooths this feature 
out.  These CLFs have several interesting implications.  First, their 
steep slopes imply that most halos above $10^{10}\msun$ host galaxies 
whose luminosities fall below current detection limits.  Second, the 
presence of faint galaxies in massive halos (largely as satellites) 
explains the observation of enhanced 
substructure around bright galaxies at $z=4$~\citep[][Figure 5]{lee06}.  
Finally, integrating these functions down to the current observational 
limit of $M_{1350} <= -18.0$ yields the mean observable occupancy in 
bins of halo mass, which we find grows from 0.05 to 1.56 as the halo 
mass increases by a factor of 10.  This finding supports the assumption 
of HODs that predict a lower LBG occupancy at lower halo 
masses~\citep{lee06}, but within our simulations it reflects the mass 
dependence of the CLF rather than a halo mass-dependent star formation 
duty cycle.

In summary, Figures~\ref{fig:hod} and~\ref{fig:clf} show that our
simulations are consistent with observed LBG occupancies of less than 
unity, but---given our near-unity star formation duty cycles---they 
predict that the ``dark" halos preferentially host galaxies that are 
simply too faint to be observed rather than galaxies that are 
temporarily dusty or quiescent.  Of course, scatter in the dust column 
and SFR at a given halo mass must contribute to the observed occupancy 
at some level, but our simulations suggest that these effects are not 
dominant.  It is possible that other feedback processes such as local
ionizing backgrounds~\citep{can10,gne10} or energy-driven outflows may 
be similarly successful in reproducing the observed occupancy, 
although momentum-driven outflows are attractive because they have already 
been shown to reproduce a wide variety of complementary constraints
(Section~\ref{sec:sims}).  The important point is that there is no 
conflict between the observed low occupancies and the smoothly-rising 
SFH scenario.  This interpretation makes several predictions, which we
discuss in the next section.

\subsection{Predictions} \label{ssec:hodpredictions}

First, invoking strong outflows in order to explain the suppressed
observed occupancy predicts that the minimum mass halo that can 
contribute to the observed sample $M_{\mathrm{min}}$ is larger than 
would be expected given the galaxy's baryonic mass and the global 
baryon fraction.  To
illustrate this, we have determined the predicted relationship 
between $M_{1350}$ and baryonic mass at $z=5.5$, multiplied by 
$\Omega_M/\Omega_b$ to obtain the minimum mass halo that could host a
galaxy with $M_{1350}$=-18 and -19, and marked these minimum halo 
masses with arrows in the top panel Figure~\ref{fig:hod}.  The actual 
minimum halo mass is roughly twice as large as would be expected 
without strong feedback~\citep[for example,][]{ber03,zhe05}.  Note that 
increasing our outflow strengths in order to improve the agreement 
with the observed colours and UV LF (Figures\ref{fig:lfz9876} 
and~\ref{fig:colcol}) would widen this gap.

Second, mass-dependent outflows that ``round off" the HOD at low
masses tend to increase the minimum mass halo that hosts an
average of one satellite galaxy, $M_1$, even more than they 
increase $M_{\mathrm{min}}$.  Consequently, we find that the ratio 
$M_1/M_{\mathrm{min}}$ is 17--18 for both simulated LBG samples.
~~\citet{zhe05} noted that simulations that do not treat outflows 
predict a ratio of 14 whereas a SAM that did predicted ratios of 
roughly 18.  By contrast, low redshift observations indicate a ratio 
of 18~\citep{zeh05,zhe07}, in excellent agreement with our outflow 
model.  These considerations suggest that galaxy clustering 
measurements are sensitive to the effects of galactic outflows,
although further inquiry is required in order to understand how
constraining these measurements are.

Third, if the galaxies living in ``dark" halos are LBG ``cousins" 
that happen to be quiescent or dusty, then the complementary galaxy 
fraction should be visible in optically-selected samples.  Hence the
smoothly-rising SFH scenario predicts that UV-selected and 
optically-selected samples should largely overlap.~\citet{bra07} 
used a Balmer break selection to identify samples of galaxies at 
$z\sim2.4$ and 3.7 that directly test this idea.  They found that, 
while some of their $z\sim2.4$ sample possess red UV continua indicative 
of significant dust or a less active star formation phase, the majority 
of their $z\sim3.7$ sample possess blue continua and would satisfy LBG 
colour cuts even though they were selected on the basis of their Balmer 
breaks.  This finding argues against the existence of a significant 
population of quiescent or dusty B-dropout ``cousins" (see 
also~\citealt{ove09}).  Note that this conclusion is not in conflict 
with the observation that $\geq80\%$ of massive 
($M_*>5\times10^{10}\msun$) galaxies at $z\geq3.5$ may be too red for 
LBG selections~\citep{man09} because the majority of star-forming 
galaxies at $z\geq4$ are not expected to be this massive--indeed,
observations suggest that galaxies become increasingly low-mass and 
dust-free at redshifts above 
3~\citep[for example,][]{bra07,ver07,bou09a,bou10a}.  Further exploration of 
optically-selected samples at $z\geq4$ would be useful in confirming 
the results of~\citet{bra07} as their samples were rather small 
(23 candidates at $z\sim4$) and lacked spectroscopic redshifts.

Fourth, the scatter in the SFR-$M_*$ relation should be small since 
short duty cycles would give rise to a significant number of galaxies 
that were transitioning between active and quiescent states.  
For example, assuming that galaxies transition between active and 
quiescent states on a halo dynamical time of $\approx100$ Myr and that 
the active star-formation period lasts 400 Myr~\citep{lee09}, roughly 
half the number of actively star-forming LBGs should be observable in a 
transition state below the star-forming ``main sequence".~\citet{lab09} 
measure a small scatter of $0.25$ dex at $z=7$ and do not observe a 
transition population even though their observations are sensitive to 
galaxies well below the star-forming main sequence (for example, see 
the inset panel in their Figure 2).  These observations are difficult 
to reconcile with short star formation duty cycles.

Finally, we predict that the observed occupancy should not be lower 
for fainter samples than for brighter ones because a larger fraction of 
halos host faint galaxies than bright galaxies (Figure~\ref{fig:clf}). 
This is visible in the bottom panel of Figure~\ref{fig:hod}, where thin
and thick curves indicate the cumulative occupancy without and with the
contribution of satellite galaxies.  If satellite galaxies are ignored
then the occupancy, defined in this case as the fraction of halos above
$M_{\mathrm{min}}$ that host at least one observable galaxy, is roughly
the same for fainter and brighter samples.  If satellite galaxies are
included then the occupancy, now defined as 
$N_g/N_h (M_h>M_{\mathrm{min}})$, 
increases for fainter samples owing to the contribution of satellites.
Note that this prediction is in direct conflict with a scenario in 
which low-mass galaxies at high redshift possess bursty SFHs 
reminiscent of local dwarf galaxies.  Unfortunately,~\citet{lee09} 
assumed a single duty cycle and did not explore this question in
detail.  Intriguingly, however, their brightest subsample at $z\sim4$ 
prefers duty cycles of 10--50\% whereas their full sample---whose
clustering properties are presumably dominated by the more numerous
faint galaxies---prefers longer duty cycles of 25--100\% (their 
Figure 9), in qualitative agreement with our prediction.  More
detailed exploration of the luminosity dependence of the occupancy
should be possible with current and upcoming data sets and would 
constitute a useful test of our predicted SFHs.

\section{Trends at Lower Redshifts} \label{sec:lowz}
Observations at $z<6$ indicate that galaxies eventually depart from the
smoothly-rising, scale-invariant SFHs that are observed at earlier
epochs.  In this Section, we comment on these observations and discuss 
relevant physical processes that could emerge at $z<6$.

We begin by considering whether star formation is a steady-state 
process at all masses.  The SFHs of actively star-forming galaxies 
more massive than $10^9\msun$ are likely to be predominantly smooth at 
all redshifts.  Observations in support of this view include the 
tight observed scatter in the SFR-$M_*$ relation ($<0.3$ dex) down 
to $z=0$~\citep{noe07a,bri04} and the fact that visibly merging 
galaxies account for less than $30\%$ of all star formation during 
$z\sim$0.24--0.80 (\citealt{jog09}; see also~\citealt{wol05}).  

At lower masses, evidence that dwarf galaxies experience short-lived 
starbursts has historically prompted many to favor generally bursty 
SFHs for dwarfs~\citep{sea72}.  Starbursts could owe to interactions 
or to an irregular gas inflow rate.  However, more recent studies 
employing larger samples have thrown the significance of starbursts in 
dwarfs' SFHs into doubt.  For example,~\citet{lee09} found that dwarfs 
whose emission lines possess high equivalent widths live preferentially 
in underdense regions.  This is inconsistent with the view that 
interaction-induced starbursts dominate dwarfs' SFHs because interactions 
are more frequent in overdense regions.  It suggests instead that star 
formation is suppressed in dwarfs that become satellite galaxies.  This 
is expected if inflows drive star formation because inflows funnel gas 
preferentially onto the central galaxy.  Note that we expect inflows 
to be smooth at $z>6$ because most observed galaxies are central 
galaxies (compare thin and thick curves from the bottom panel of 
Figure~\ref{fig:hod}).  

More recently,~\citet{jlee09} found that only 
6\% of a volume-complete sample of $\approx300$ local dwarf galaxies 
are currently undergoing starbursts.  Further, bursts accounted for 
at most one quarter of the total current star formation in their sample.  
Together with~\citet{jog09}, these results indicate that star formation 
in the central galaxies of halos below $10^{12}\msun$ is a 
predominantly smooth process at all redshifts, which is a prediction of 
our model as well as others~\citep[for example,][]{guo08,cat10}.

Departures from rising SFHs are observed after $z=2$ in two senses.  
First, the emergence of a bimodality in galaxy colours as early as 
$z=2$~\citep{gia05} reflects the fact that many galaxies eventually cease 
forming stars.  These ``quenched" galaxies live preferentially in overdense 
regions~\citep{hog04,coo10}, which supports the view that quenching owes 
to the onset of hot-mode accretion in halos more massive than 
$\approx10^{12}\msun$~\citep{ker05,ker09}, possibly coupled with extra 
heating from a central source~\citep{cro06,cat06}.  These effects should 
be weak in current $z>6$ samples because their host halos are generally 
less massive than $10^{12}\msun$ (Figure~\ref{fig:hod}).

Second, observations indicate that the SFRs of galaxies whose SFHs 
are not yet quenched decrease for $z<1$~\citep{noe07a,pe08}.  This 
cannot be attributed to evolution in the interaction rates because 
observations indicate that the merger rate evolves either 
mildly~\citep{cas05} or not at all~\citep{lot08} out to $z\sim1$.  
On the other hand, it occurs naturally if the gas accretion rate is 
assumed to trace the host halo's growth rate because halo growth 
rates fall off at late times owing to cosmological expansion.  Models 
that incorporate this effect readily yield rising SFHs at early 
times and decaying SFHs at late times~\citep{dav08,bou09,dut09}.

Finally, observations indicate that the slope of the SFR-$\mstar$ 
relation flattens from $\sim$unity at $z=6$ to 0.7--0.9 at 
$z<2$~\citep[for example,][]{bri04}.  This departure from scale-invariance 
is equivalent to downsizing~\citep{cow96}.  Our simulations predict a 
slope of $\approx0.9$ at $z\leq2$~\citep{dav08}, hence they qualitatively
reproduce the observed evolution.  Two effects contribute.  First, the 
fraction of galaxies living in massive ($>10^{12}\msun$) halos increases 
with time and mass.  If the early stages of quenching (see above) involve 
suppressed inflow rates and a gradual exhaustion of available gas 
reservoirs, then the amount of suppression should increase with time and 
mass.  This should manifest both as an overall flattening of the SFR-$\mstar$ 
relation~\citep{dav08}, and as a dependence of the slope on environment.  
Second, feedback processes that delay gas accretion or star formation in 
low-mass halos flatten the predicted slope~\citep{noe07b,bou09}.  
Momentum-driven outflows are a natural candidate because the 
mass-loading factors scale inversely with the velocity dispersion.
This in turn causes gas in low-mass halos to spend most of its time 
suspended above the galaxy's ISM.  Outflows are also appealing because 
they reconcile our simulations with a  wide range of complementary 
observations (Section~\ref{sec:intro}).~~\citet{dut09} argued that 
outflows cannot change the slope of the predicted SSFR trend, but their 
model did not treat the possibility of outflows that do not escape from 
the halo.  

While our simulations qualitatively reproduce the observed flattening 
trend, in detail they predict slightly less flattening than is observed.  
This could suggest either that inflows in massive halos are 
undersuppressed~\citep{gab10}, or that star formation in low-mass halos is delayed by 
other processes in addition to outflows.  One possibility is heating by 
the local ionizing background within each galaxy~\citep{can10,gne10}.  
Our model does not treat the local background, but it would be 
interesting to explore in future work.

In summary, star formation in halos below $10^{12}\msun$ is likely to
be a smooth process at all redshifts, but below $z\approx$2--3 the SFRs 
begin to decline and the SFR-$\mstar$ trend flattens.  The
emergence of declining SFRs following $z\approx2$ owes to the decline 
in halo growth rates, which in turn owes to cosmological expansion.
Meanwhile, the departure from scale-invariance owes to at least two 
effects.  In massive halos, accretion shocks and AGN feedback likely 
suppress inflows.  In low-mass halos, outflows expel cold, star-forming 
gas while ionizing backgrounds heat it more rapidly than it can be
replaced by fresh inflows.

\section{Summary}
\label{sec:sum}
We have conducted a detailed comparison between the predictions of
state-of-the-art cosmological hydrodynamic simulations including
momentum-driven outflows and recent observations of reionization-epoch 
galaxies in order test two fundamental predictions regarding 
high-redshift SFHs:
\begin{enumerate}
\item SFHs at early times are smoothly-rising; and
\item SFH shapes are scale-invariant.
\end{enumerate}
Our findings are as follows:
\begin{enumerate}
\item Our simulations reproduce the observed rest-frame UV luminosity
function as well as its integral to current limits to within a
factor of 2--3 during the epoch $z=$6--8.  They also reproduce the
observed stellar mass density to current limits.  Either of these
constraints could in principle be matched by any SFH at a single 
epoch, but the only smooth SFH that can account for their evolution 
is a rising one.  This suggests that our prescriptions for star 
formation and feedback yield SFHs in reasonable agreement with 
current constraints.  In detail, we find that increasing the assumed 
outflow mass-loading factors by $\approx$50\% could bring the 
predicted and observed UV luminosity densities into better agreement 
without compromising the good agreement with the observed stellar 
mass density (to within the reported errors).  Integrating below 
current detection limits, we estimate that current observations 
probe the brightest 30\% of the total star formation and stellar 
mass densities at $z\geq6$. 
\item We reproduce the observed blue UV continua at $z=7$ for all
but the faintest galaxies and marginally reproduce the observed 
bluening to lower luminosities.  The UV continua of the faintest 
galaxies may suggest lower metallicities and a nonzero ionizing 
escape fraction~\citep{bou10a}.  We attribute the observed 
colour-magnitude trend to a modest level of dust reddening that 
increases with luminosity as seen at lower 
redshifts~\citep{meu99,sha01}.  This suggests that exotic stellar 
populations are not required in order to explain observations of
galaxies brighter than $H_{160}=27.5$ at $z=7$~\citep{fink09,bou10a}.
\item Our simulations' relatively evolved stellar populations
predict red $H_{160}-[3.6]$ colours at $z\geq6$, augmented by dust and 
optical emission lines.  However, the predicted  colours are 
$\approx0.5$ mag bluer than reported by~\citet{lab09}.  It is not 
clear whether resolving this discrepancy requires significantly higher 
mass resolution or improved observations.  For the brightest galaxies, 
allowing the redshift and extinction to vary freely enables our models 
to reproduce the observed $K$--$[4.5]$ fluxes while underproducing the 
observed $J_{125}$ and $H_{160}$ fluxes.  This suggests that the red 
observed $H_{160}-[3.6]$ colours may be as likely to owe to systematic 
error in the WFC3 as in the IRAC observations.
\item We reproduce the near-unity slope of the observed SFR-$\mstar$ 
relation at all redshifts.  This agreement supports a scenario in which
observed reionization-epoch galaxies began forming stars at similar
epochs and possess SFHs that vary only by a scale factor, which is a 
robust prediction of hydrodynamic simulations.  We also reproduce 
the observed slow evolution of the SSFR, which is incompatible with
constant or smoothly declining SFHs while supporting
smoothly-rising SFHs~\citep{sta09}.  The predicted scatter is 
$\approx 0.1$dex, less than the (still remarkably tight) reported 
scatter of 0.25 dex.  The larger observed scatter may imply a need 
for higher mass resolution in order to resolve the impact of minor 
mergers, but it may also be boosted by observational uncertainties.  
An offset of a factor of 2--3 between observations and predictions 
suggests that observationally-inferred stellar masses should be 
corrected for the likely impact of optical emission lines.
\item Our significantly improved treatment for the production and
transport of metals predicts that observable galaxies at $z\geq6$ 
possess metallicities in excess of 0.1$Z_\odot$.  This is several 
times larger than our previous prediction.  Increasing the assumed 
outflow mass-loading factors to improve the agreement between the 
predicted and observed UV LFs would lower these predictions by 
$\sim0.1$ dex, but it would not change the fundamental result that 
galaxies self-enrich quite quickly.  Hence we confirm that it 
should be possible to understand current samples without reference 
to Population III stars.  Nevertheless, a residual fraction $10^{-3}$
of all star formation at $z=6$--8 may occur below the Population
III metallicity threshold, which may eventually be visible as 
pair-production supernovae or GRBs.
\item There is no conflict between observations suggesting that many 
massive halos do not contribute to the observed samples at $z=4$--5 
and the smoothly-rising SFH scenario.  This is because clustering 
observations constrain the halo occupancy rather than the star 
formation duty cycle.  The latter is only one of several possible 
interpretations of an overall occupancy that is less than unity, 
and our simulations disfavor it.  Instead, we favor the interpretation 
that it reflects the scatter in the luminosity-halo mass relation.  
This is because strong outflows boost the scatter in the SFRs of 
low-mass halos, suppressing the UV luminosities of many isolated 
central galaxies that would be visible in the absence of outflows.  
Consequently, more than half of observed galaxies live in halos whose 
mean (observable) occupancy is less than one and the overall fraction 
of halos above $M_{\mathrm{min}}$ that host an observable galaxy is 
0.2--0.3 as observed even though the predicted star formation duty 
cycle is unity.  A halo occupancy in this range should thus be 
regarded as indicative of a high star formation duty cycle rather 
than a low one.  This interpretation makes a number of predictions:
\begin{enumerate}
\item The minimum mass for hosting an observed galaxy $M_{\mathrm{min}}$ 
is at least twice as large as would be expected given galaxies' baryonic 
masses and the global baryon mass fraction;
\item The minimum mass for hosting an average of two galaxies is larger
than $M_{\mathrm{min}}$ by a factor of 17--18;
\item Optically-selected and UV-selected samples should largely overlap
at $z\geq4$;
\item The scatter in the SFR-$\mstar$ relationship should be small, and
few galaxies should be observed with suppressed SSFRs indicative of a
transition between active and quiescent states;
\item The overall occupancy for faint samples should be greater than or 
equal to the overall occupancy for brighter samples, in direct conflict
with a picture in which lower-mass galaxies possess burstier SFHs in
analogy with local dwarf galaxies.
\end{enumerate}
Each of these predictions is in tentative agreement with current 
observations, but we expect that more detailed analyses of current and
upcoming samples will test them more thoroughly.

\end{enumerate}

Further progress in understanding high-redshift SFHs will occur on both 
theoretical and observational fronts.  On the theoretical side, it would 
be interesting to increase our mass resolution significantly while 
relaxing the assumption that wind particles temporarily decouple 
hydrodynamically in order to resolve feedback on small scales.  
Simulations of dwarf galaxy formation at low redshift tend to find 
burstier SFHs as the mass resolution is 
increased~\citep[for example,][]{val08} owing to stochastic
processes~\citep{ger80}, hence it is natural to suppose that similarly
low-mass objects at $z\geq6$ would also be bursty.  On the other hand,
dwarf galaxy SFRs are generally $<0.5 \msun\mbox{yr}^{-1}$ whereas even 
the faintest $z=7$ galaxies indicate SFRs of $>2\msun\mbox{yr}^{-1}$; 
whether bursty SFRs are expected in the presence of efficient gas 
inflows is unclear.  If so, this may improve the agreement with the 
observed scatter in the SFR-$\mstar$ relation as well as the colours of 
low-mass galaxies~\citep{lab09}.  Unfortunately, increasing our 
mass resolution is not trivial 
because our present simulations already probe current computational 
limitations.  Further progress will require resimulations of a 
statistical sample of high-redshift galaxies~\citep[for example,][]{bro07}.  
By contrast, incorporating stronger outflows in order to improve the 
agreement with the observed UV luminosity function would be trivial 
(aside from the computational expense), although it remains to be seen 
whether this would also preserve the agreement with the observed density 
of metals in the intergalactic medium~\citep{opp09b}.  

The impact of an inhomogeneous ionizing background warrants closer 
scrutiny.  Our current work does not indicate that our assumption of a 
uniform~\citet{haa01} ionizing background produces large errors in our 
predictions (DFO06).  Nonetheless, it would be useful to study 
the relative roles of photoionization heating versus outflows in 
suppressing star formation in faint galaxies using simulations that 
treat the nascent ionizing background self-consistently.  Preliminary
work suggests that interesting nonlinear interactions exist between 
outflows and the ionizing background; these warrant more detailed 
study~\citep{paw09}.

A closer inquiry into the nature of our predicted HOD and its ability 
to reproduce clustering constraints at $z=4$--5 is in order.  We have 
not delved into a full clustering analysis because 
our goal has been to demonstrate that clustering properties in the 
presence of mass-dependent outflows are qualitatively different than 
the conventional central-satellite HOD picture and have the potential 
to reconcile the observed duty cycle with smoothly-rising SFHs.  To 
this end, we have already demonstrated encouraging agreement between 
our predictions and the observed occupancy and $M_1/M_{\mathrm{min}}$.  
A more detailed study of the CLF and its associated HOD in simulations 
with and without outflows would bring the impact of outflows on the 
HOD into sharper relief while providing a new class of template HODs 
for interpreting observations.  Such a study would involve using our 
predicted HOD to populate halos from a larger-volume N-body simulation 
in order to provide good statistics, a task that is well beyond the 
scope of our current work.  

On the observational side, further progress will follow from a more 
systematic reconstruction of the high-redshift SFR-$\mstar$ correlation 
and its evolution, which are the most important observational probes 
of the ``typical" SFH in an epoch where the SFHs of individual 
galaxies remain unconstrained.  This involves work on three fronts:
First, deeper rest-frame optical constraints on larger samples of 
galaxies are required in order to leverage existing UV-selected 
samples.  Second, progress requires improved SED modeling that invokes 
physically-motivated SFHs (rather than, for example, instantaneous 
bursts) and accounts for nebular emission in order to reduce 
systematic errors.  Third, uniform SED-fitting studies incorporating 
common modeling assumptions~\citep[for example,][]{sch10} will 
reduce artificial scatter owing to differing modeling assumptions.  
Because of the importance of the SFR-$\mstar$ plot, we heartily 
encourage the publication of inferred physical properties
order to facilitate further theoretical comparisons.

In order to gain a better understanding of the scatter in the 
SFR-$\mstar$ relation as well as the implications of the low observed 
halo occupancy, we encourage further inquiry into the overlap between 
UV-selected and optically-selected samples at $z\geq4$, following the 
example of~\citet{bra07}.  These studies are a strong test of our
prediction that SFHs are predominantly smooth rather than bursty.

With the growing body of observational evidence that high-redshift 
SFHs are smooth rather than bursty and the implication that the 
progenitors of $V$-dropouts, for example, are visible as 
$i$-dropouts, it is important that SED-fitting studies account for 
galaxies' hypothetical progenitors at earlier epochs in addition
to their observed fluxes.  For example, exponentially decaying SFHs
with a decay time of 300 Myr applied to $V$-dropouts predict 
bright progenitor $i$-dropouts whose colours/SSFRs are observationally 
ruled out (unless they are very dusty, which again reinforces the 
need for study of optically-selected samples).  For the same reason,
it is now becoming possible to constrain high-redshift galaxy SFHs 
using statistical methods applied at multiple 
epochs~\citep[see, for example,][]{con09,mar10,pap10}.  Such studies 
provide complementary tests of our predicted SFHs.

Finally, we have outlined a number of predictions regarding the 
clustering properties of faint ($M_{1350}>-20$) high-redshift 
galaxies that may be amenable to study using current samples
(for example, from GOODS) and that should be considered in the 
design of upcoming Hubble and Spitzer programs to study the 
distant universe.

\section*{Acknowledgements}
We thank Nicolas Bouch\'e, Rychard Bouwens, Dawn Erb, Steve Finkelstein, 
Valentino Gonzalez, Alaina Henry, Ivo Labb\'e, Peng Oh, Casey Papovich,
Moire Prescott, Naveen Reddy, Michele Trenti and Risa Wechsler for 
suggestions and stimulating conversations.  We thank the anonymous
referee for suggestions that improved the draft.  Our simulations 
were run on the University of Arizona's Xeon cluster.  
Support for this work was provided by the NASA Astrophysics Theory 
Program through grant NNG06GH98G, as well as through grant number 
HST-AR-10647 from the SPACE TELESCOPE SCIENCE INSTITUTE, which is 
operated by AURA, Inc. under NASA contract NAS5-26555.  Support for 
this work, part of the Spitzer Space Telescope Theoretical Research 
Program, was also provided by NASA through a contract issued by the 
Jet Propulsion Laboratory, California Institute of Technology under 
a contract with NASA.  KMF gratefully acknowledges support from NASA 
through Hubble Fellowship grant HF-51254.01 awarded by the Space 
Telescope Science Institute, which is operated by the Association 
of Universities for Research in Astronomy, Inc., for NASA, under 
contract NAS 5-26555.


\end{document}